\documentclass[sigconf]{acmart}

\AtBeginDocument{%
  }

\setcopyright{acmlicensed}
\copyrightyear{2018}
\acmYear{2018}
\acmDOI{XXXXXXX.XXXXXXX}

\acmConference[Chinese CHI 2024]{Chinese CHI 2024}{Nov 22--25,
  2024}{Shenzhen, China}
\acmISBN{978-1-4503-XXXX-X/18/06}

\acmSubmissionID{79}

\usepackage{mathptmx}                  
\usepackage{paralist}
\usepackage{tikz}
\usepackage{cleveref}
\usepackage{siunitx}
\usepackage{subcaption}
\usepackage{threeparttable}
\usepackage{tabularx}
\usepackage{multirow}

\newcommand{\prefix}{\textcolor[RGB]{0, 102, 204}}
\newcommand{\indicatorred}{\textcolor[RGB]{228, 91, 123}}
\newcommand*\transparentcircled[1]{\tikz[baseline=(char.base)]{
            \node[shape=circle,draw=black,fill=none,inner sep=0.5pt, line width=0.5pt, text=black, font=\footnotesize] (char) {#1};}}
\newcommand*\darkcircled[1]{\tikz[baseline=(char.base)]{
            \node[shape=circle,draw=blue!10!gray,fill=blue!10!gray,inner sep=0.5pt, line width=0.5pt, text=white, font=\footnotesize] (char) {#1};}}
\newcommand*\casebox[1]{\tikz[baseline=(char.base)]{
            \node[shape=rectangle,draw=white!30!black,fill=gray!15!white,inner sep=1pt, line width=0.5pt, text=black, font=\footnotesize] (char) {#1};}}
\newcommand*\specialcircled[2]{
\tikz[baseline=(char.base)]{
    \node[shape=circle,draw=black,align=center,inner sep=0.5pt, font=\footnotesize, line width=0.5pt] (char) {%
        \scalebox{0.8}{#1}\scalebox{0.6}{#2}};}}
        
\begin{document}

\title{Medillustrator: Improving Retrospective Learning in Physicians' Continuous Medical Education via Multimodal Diagnostic Data Alignment and Representation}

\author{Yuansong Xu}
\orcid{0009-0005-1630-6279}
\affiliation{%
  \institution{ShanghaiTech University}
  \city{Shanghai}
  \country{China}
}
\email{xuys2023@shanghaitech.edu.cn}

\author{Jiahe Dong}
\orcid{0009-0003-1537-0717}
\affiliation{%
  \institution{ShanghaiTech University}
  \city{Shanghai}
  \country{China}
}
\email{dongjh@shanghaitech.edu.cn}

\author{Yijie Fan}
\affiliation{%
  \institution{Nanyang Technological University}
  \city{Singapore}
  \country{Singapore}
}
\email{yfan013@e.ntu.edu.sg}

\author{Yuheng Shao}
\orcid{0009-0008-6991-6427}
\affiliation{%
  \institution{ShanghaiTech University}
  \city{Shanghai}
  \country{China}
}
\email{shaoyh2024@shanghaitech.edu.cn}

\author{Chang Jiang}
\affiliation{%
  \institution{Shanghai Clinical Research and Trial Center}
  \city{Shanghai}
  \country{China}
}
\email{cjiang_fdu@yeah.net}

\author{Lixia Jin}
\affiliation{%
  \institution{Zhongshan Hospital Fudan University}
  \city{Shanghai}
  \country{China}
}
\email{jin.lixia@zs-hospital.sh.cn}

\author{Yuanwu Cao}
\affiliation{%
  \institution{Zhongshan Hospital Fudan University}
  \city{Shanghai}
  \country{China}
}
\email{cao.yuanwu@zs-hospital.sh.cn}

\author{Quan Li}
\authornote{The corresponding author.}
\affiliation{%
  \institution{ShanghaiTech University}
  \city{Shanghai}
  \country{China}
}
\email{liquan@shanghaitech.edu.cn}
\renewcommand{\shortauthors}{Xu et al.}

\begin{abstract}
  Continuous Medical Education (CME) plays a vital role in physicians' ongoing professional development. Beyond immediate diagnoses, physicians utilize multimodal diagnostic data for retrospective learning, engaging in self-directed analysis and collaborative discussions with peers. However, learning from such data effectively poses challenges for novice physicians, including screening and identifying valuable research cases, achieving fine-grained alignment and representation of multimodal data at the semantic level, and conducting comprehensive contextual analysis aided by reference data. To tackle these challenges, we introduce \textit{Medillustrator}, a visual analytics system crafted to facilitate novice physicians' retrospective learning. Our structured approach enables novice physicians to explore and review research cases at an overview level and analyze specific cases with consistent alignment of multimodal and reference data. Furthermore, physicians can record and review analyzed results to facilitate further retrospection. The efficacy of \textit{Medillustrator} in enhancing physicians' retrospective learning processes is demonstrated through a comprehensive case study and a controlled in-lab between-subject user study.
\end{abstract}




\maketitle

\section{Introduction}
\par Physicians utilize diagnostic records for continuous self-learning and skill enhancement beyond immediate patient care~\cite{10.1371/journal.pone.0288474}. This ongoing engagement is crucial for refining diagnostic practices and improving skills~\cite{Shimizu2023}. Continuous Medical Education (CME)~\cite{Marinopoulos2007EffectivenessOC,Bordage2009ContinuingME} complements this effort, encouraging physicians to engage in self-directed learning through meticulous case studies and collaborative discussions with experienced peers. To support physicians' CME, especially novices, existing literature explores pedagogical theories like Self-Directed Learning (SDL)~\cite{knowles1975self,liu2021story,tagawa2008physician} and Case-Based Learning (CBL)~\cite{mclean2016case}. These theories stress self-motivated improvement and the integration of practical, case-based learning in medicine. Guided by these theories, technical solutions, including learning platforms and interactive systems~\cite{Raidou2018BladderRV,Mrth2020RadExIV,Yang2023LeveragingHM}, have been developed to enhance novice physicians' learning experiences. While many platforms use clinical resources like online materials~\cite{online_videos}, slides, and videos~\cite{Medscape,VisualDx}, interactive systems excel by incorporating diverse multimodal diagnostic data~\cite{Raidou2018BladderRV,Mrth2020RadExIV}. This data encompasses radiology images (e.g., X-rays, CT scans, MRI), clinical texts (e.g., diagnostic reports), and laboratory indicators, significantly improving physicians' retrospective learning~\cite{cheng2021vbridge,Yang2023LeveragingHM}.

\par While employing multimodal diagnostic data to enhance the retrospective learning experience of physicians has demonstrated effectiveness, several challenges persist. First, a significant challenge is \textbf{\textit{the lack of efficient methods for screening and identifying high-value research cases.}} Novice physicians, despite theoretical trained, face difficulties in matching symptom descriptions with real patient cases~\cite{BMCPrimaryCare}. Clinical practice reveals complex and variable patient conditions, where reliance on ``textbook-style'' identification is inadequate~\cite{Donner-Banzhoff353,Bridgingthegap}. Patients with similar images and lab results may receive different diagnoses due to subtle disparities in other diagnostic data~\cite{Impactofdiagnostic}. These cases, requiring thorough analysis, are valuable for retrospective learning. However, effectively screening and discovering them is challenging, akin to finding a needle in a haystack for novice physicians. Senior physicians rely on experience but face constraints due to limited exposure to valuable cases. Thus, there's a need to identify high-value cases strategically, enabling efficient learning and continuous improvement in medical practice. Second, \textbf{\textit{retrospective learning requires precise alignment and effective representation of data at the semantic level.}} Novice physicians, particularly those early in their careers, struggle with interpreting image data due to its complexity compared to text descriptions and indicators~\cite{doi:10.2214/AJR.19.21802}. For instance, identifying subtle lines or changes indicating fractures on CT images can be challenging, leading to potential confusion with other bone lesions despite explicit mentions in diagnostic reports. This underscores the need for enhanced presentation of image data to facilitate comprehensive analysis, possibly by integrating other modal data for reference. However, existing research on multimodal data analysis primarily focuses on prediction and decision-making~\cite{Tang2021VideoModeratorAR, Wu2023LiveRetroVA, Zeng2022GestureLensVA}, or exploring joint effects between modalities~\cite{Liang2022MultiVizTV, Wang2021M2LensVA}, with alignment often limited to specific dimensions like time or individuals for particular segments. While tools for cross-modal medical data analysis exist, such as automated image segmentation~\cite{Wang2022PyMICAD, Wang2020AnnotationefficientDL} and medical phrase grounding~\cite{Chen2023MedicalPG, Li2023ACS}, they typically analyze images only at the object detection level, lacking finer semantic alignment. Additionally, current commercial tools often present different data modalities separately, requiring physicians to switch between sources like images, diagnostic reports, and indicators during case study discussions. This inconvenience not only complicates the learning process for presenters but also imposes cognitive load on listeners. Therefore, there's an urgent need to integrate relevant multimodal data into a unified and coherent representation format.
Third, \textbf{\textit{ensuring the establishment of contextual references becomes imperative for interpretation and communication during the learning process, even when data is appropriately aligned and represented.}} Experienced physicians possess an innate ability to incorporate comparisons and references into their subjective analysis of diagnostic data. Drawing on their extensive clinical experience, they instinctively identify abnormalities by contrasting them with normal data~\cite{Usingcognitivetheory}. This experiential knowledge, although subtle and implicit, significantly contributes to their diagnostic prowess. However, current research often focuses on analyzing specific diagnostic data without incorporating the normal range as a context for comparison~\cite{Silberberg2021Melding}. This presents a significant challenge for novice physicians who lack sufficient experience to discern abnormalities effectively. Cheng et al.~\cite{cheng2021vbridge} further underscored the importance of reference values, highlighting that ``\textit{the reference values are vital in facilitating prediction interpretations for clinicians}'' in both static and dynamic contexts.

\par In this study, we focus on enhancing physicians' retrospective learning through integrating multimodal diagnostic data, targeting continuous medical education. Through a formative study involving six physicians, we identified current patterns and concerns in retrospective analysis and derived design requirements to address them. Following Bruner's discovery learning theory~\cite{Bruner1960ThePO}, we structured these requirements into three levels: \textbf{(1)} leveraging prior knowledge for broad concept understanding (\textbf{overview}), \textbf{(2)} promoting active engagement and inquiry-based learning (\textbf{detail}), and \textbf{(3)} providing scaffolding for comprehension through various representations (\textbf{retrospection}). Our solution, \textit{Medillustrator}, is a visual analytics system tailored to aid novice physicians in retrospectively analyzing multimodal diagnostic data. It offers a structured approach, allowing users to begin with an overview for screening research cases, delve into specific cases with multimodal data and references, and record analyzed results for review and retrospection. To validate its effectiveness and usability, we conducted a comprehensive case study with a controlled user study. In summary, the main contributions of this study are as follows:
\begin{compactitem}
\item We identify current patterns and challenges in physicians' retrospective analysis through the formative study.
\item We develop a visual analytics system with a structured data processing and modeling pipeline to present consistent multimodal diagnostic data alignment and visualization to facilitate the learning experiences of novice physicians.
\item We demonstrate the effectiveness and usability of our approach through a case study and a controlled user study.
\end{compactitem}

\section{Related Work}
\subsection{Continuous Medical Education}
\par Beyond traditional medical education, continuous medical education (CME)~\cite{Marinopoulos2007EffectivenessOC,Bordage2009ContinuingME} is crucial in the lifelong learning of physicians. This ongoing educational process is essential for health professionals to consistently uphold, update, and refine their knowledge, skills, and attitudes, thereby ensuring proficient practice. CME can be categorized into three distinct types based on learning formats: \textit{active learning}, \textit{passive learning}, and \textit{unstructured learning}~\cite{31757278,stephenson2020}. \textit{Active learning} includes structured training activities like seminars, workshops, and conferences, where participants actively engage in the learning process~\cite{2021nguyen}. \textit{Passive learning} emphasizes the acquisition of knowledge from external sources such as medical journals, online resources, and other self-study materials. Meanwhile, \textit{unstructured learning} focuses on self-directed exploration within practical work settings, such as clinical practice, case studies, and collaborative discussions~\cite{2023donkin}.
\par This study concentrates on enhancing the retrospective learning experience of novice physicians by examining clinical diagnostic data. This involves engaging in various learning formats, such as participating in workshop discussions for \textit{active learning}, analyzing existing materials for \textit{passive learning}, and independently exploring data from diverse sources in \textit{unstructured learning}.

\subsection{Diagnostic Multimodal Data Analysis}
\par Analyzing clinical diagnostic data from a single patient often yields information in multiple modalities, such as radiology images, medical texts, and laboratory test results. However, conducting a comprehensive analysis during the data analysis and learning process poses a significant challenge~\cite{Cui2022DeepMF}. To address this challenge, deep learning-based methods have been developed for multimodal data analysis, employing multimodal fusion to extract and model complex relationships across different modalities~\cite{Wang2021ModelingUI,Holste2021EndtoEndLO,Lu2020AIbasedPP,Duanmu2020PredictionOP}.
\par Existing work in this domain can be categorized based on two fusion strategies: \textit{decision-level fusion} and \textit{feature-level fusion}~\cite{Huang2020FusionOM}. \textit{Decision-level fusion} involves independently obtaining prediction results for each modality and subsequently fusing these predictions. This approach allows for fusing different modalities without retraining the unimodal models, offering flexibility and simplicity. Fusion strategies employed in existing work include \textit{majority vote}, \textit{weighted sum}, and \textit{averaging}~\cite{Rokach2010EnsemblebasedC,Zhu2022MedicalLS,Shehanaz2021OptimumWM}. Specifically, Wang et al.~\cite{Wang2021ModelingUI} proposed a joint multimodal fusion algorithm that considers model uncertainty when estimating correlations among predictions from different modalities. Holste et al.~\cite{Holste2021EndtoEndLO} utilized the output probabilities of unimodal predictions to fuse image-derived features with tabular non-image features, while Huang et al.~\cite{Huang2020FusionOM} introduced a meta-neural network classifier trained with predicted probabilities from both medical images and electronic medical records (EMR). However, \textit{decision-level fusion} integrates information only at the prediction stage, overlooking interactions between underlying features. This approach limits information integration for the relationships across multiple modalities.

\par On the other hand, \textit{feature-level fusion} involves combining original data and extracted features from different modalities into multimodal hidden representations for final decision-making, offering advantages in capturing intricate relationships between modalities~\cite{Cui2022DeepMF}. For example, Lu et al.~\cite{Lu2020AIbasedPP} concatenated clinical features with learned pathology image features for classifying primary or metastatic tumors and determining origin sites. Duanm et al.~\cite{Duanmu2020PredictionOP} utilized the learned feature vector of non-image modality along with image features at multiple layers to predict responses to chemotherapy in breast cancer. The literature also explored connections and joint effects among different modalities. In a study by Chen et al.~\cite{Chen2023MedicalPG}, the focus lies on the medical phrase grounding (MPG) task, presenting strategies for contextually aligning text descriptions with corresponding regions of interest (ROIs) in medical images. Another study by Qin et al.~\cite{Qin2022MedicalIU} employed vision-language models, focusing on formulating effective prompts for object detection in medical imagery. Additionally, Chen et al.~\cite{cheng2021vbridge} proposed a visual analytics system designed for analyzing features in electronic health records (EHR) through contribution-based post hoc explanations. While these studies present effectiveness, they primarily targeted diagnostic prediction and decision-making. Besides, the multimodal alignments predominantly concentrated on the object detection level with limited fine-guaranteed analysis. 

\par Our work distinguishes itself from prior research in facilitating novice physicians' retrospective learning, with concerns about multimodal data alignment through feature-level fusion at the semantic level. We also incorporate contextual information for explicit interpretation and reference, aiming to reveal the implicit diagnostic experience of senior physicians based on a comparison of reference values.

\subsection{Multimodal Medical Data Visualization}
\par Medical data not only holds critical information for treatment decisions but also plays a pivotal role in retrospective learning through data analysis. While analyzing heterogeneous medical data from various sources poses challenges, visual analytics systems, integrating interactive visualization methods with statistical inference and correlation models, have demonstrated the potential to aid users in effective analysis, thereby concealing the underlying complexity of the data~\cite{Caban2015VisualAI}.
\par Besides the works that concentrate on individual data modalities~\cite{Wang2021ThreadStatesSV,Shahar2006DistributedII,sultanum2022chartwalk}, existing studies have introduced multimodal data visualization tools to facilitate comprehensive analysis. For example, Raidou et al.~\cite{Raidou2018BladderRV} presented tools that enable detailed visual exploration and analysis of how variations in bladder shape impact the accuracy of dose delivery. Bannach et al.~\cite{Bannach2017VisualAF} combined medical image analysis with visual analytics of patient data to analyze patient cohorts.

\par While multimodal data visualizations have demonstrated effectiveness, they often align various data modalities based on specific dimensions such as patient or timestamp~\cite{cheng2021vbridge, Yang2023LeveragingHM}, resulting in separate presentations for each data modality. This segmented presentation limits the coherent analysis of patient cases, hindering a comprehensive understanding across different modalities. To enhance the coherence and unity of presentation in physicians' multimodal medical data analysis, our approach centers on image modality, complementing it with overlays of additional modalities (e.g., diagnosis text). This method emphasizes the pivotal role of image data in diagnostic analysis, facilitating a deeper and more intuitive understanding by merging visual and textual data into a unified, cohesive representation. Additionally, we integrate indicators and supplementary data, even when not directly tied to the primary information, to serve as reference points for diagnostic analysis.

\section{Formative Study}
\par To address physicians' challenges in retrospective learning, we conducted a formative study to pinpoint their specific difficulties in this phase. Analyzing the study results yielded valuable insights, informing the formulation of design requirements for our approach.

\subsection{Participants and Procedure}
\par We engaged with a team of six domain experts (\textbf{E1-E6}) (\autoref{tab:participants_info}) from a local hospital, consisting of four novice physicians (\textbf{E1}: male, \textbf{E2}: male, \textbf{E4}: male, \textbf{E6}: male) and two experienced physicians (\textbf{E3}: female, \textbf{E5}: male). While \textbf{E1} and \textbf{E3} are in the same department, the others are from different departments. Each physician brings clinical experience and actively engages in medical education activities. With approval from the Institutional Review Board (IRB), we conducted semi-structured interviews to glean insights and ideas from the physicians. Throughout these interviews, participants addressed open-ended questions about their typical practices and approaches in retrospective learning, also sharing the challenges they face in their practices.

\begin{table}[h]
\vspace{-3mm}
\small 
\centering
\caption{Participant information includes gender (F/M) and experience levels (A/B), with A denotes years of teaching experience and B means years of clinical experience.}
\vspace{-3mm}
\begin{tabular}{@{}ccccc @{}}
\toprule
ID & Gender/Age & Exp & Position & Specialty \\ \midrule
E1 & F/32 & 1/5 & Resident & Cardiology \\
E2 & M/36 & 2/9 & Attending Physician & Endocrinology \\
E3 & F/51 & 15/23 & Chief Physician & Orthopedics \\
E4 & M/32 & 0/4 & Resident & General Practice\\
E5 & M/47 & 10/18 & Associate Chief Physician & Cardiology \\
E6 & M/28 & 0/1 & Intern & Orthopedics \\
\bottomrule
\end{tabular}
\label{tab:participants_info}
\vspace{-5mm}
\end{table}


\subsection{Analysis and Results}
\par The interviews were recorded with participants' consent and transcribed into text. Thematic analysis~\cite{boyatzis1998transforming} was used to analyze the data and derive qualitative findings, due to its ability to reveal patterns and themes in qualitative data, offering insights into participants' experiences, perspectives, and behaviors~\cite{clarke2017thematic}. The challenges identified during physicians' retrospective learning are summarized below:
\par \textbf{C1. Sifting through patient data to find valuable learning cases is inefficient and time-intensive.} Novice physicians often find themselves tasked with patient diagnoses and retrospective studies under the guidance of senior physicians. Given the inherent value of physicians' time, the crucial aspect of ensuring the quality of learning lies in the analysis of high-value case data. However, sifting through tons of patient cases to identify valuable learning cases proves to be a challenging and time-consuming process. \textbf{E2} shared their experience, stating, ``\textit{I tried going through clinic cases during my internship, thinking I'd dig up some valuable insights. But, turns out, sifting [through them] manually was a real-time sink – inefficient and didn't give much insight for all the time I put in.}'' The current approach relies on senior physicians to select and summarize typical cases presented in seminars for clinical case teaching and discussion. However, this method, relying on accumulated years of experience, is equally inefficient. According to \textbf{E5}, a senior physician, ``\textit{Even with all my experience, picking cases to teach new physicians is tricky. It's not just about going for the complicated ones; it's about finding the ones that pack the most educational punch. Depending too much on personal judgment and experience might unintentionally miss out on less obvious but just instructive cases, which ends up narrowing the learning scope for [new] physicians.}''

\par \textbf{C2. Identifying areas of abnormality and pathology is challenging.} Among the various modalities of diagnostic data, physicians generally acknowledged the crucial impact of image data during diagnostic analysis. As mentioned by \textbf{E3}, ``\textit{In real-world practice, images like X-rays or MRIs give us a clear, intuitive understanding of a patient's condition. These visuals can uncover issues that are hard to spot using other methods and play a crucial role in supporting further diagnosis.}'' When physicians review medical images, the primary goal is to identify areas of interest for further analysis. These areas could include lesions or lumps in the case of detecting lung cancer, a task that remains challenging for novice physicians during the comprehensive analysis of data from other modalities. As highlighted by \textbf{E6}, ``\textit{We often find it challenging to interpret [medical] images, especially when it comes to spotting subtle issues like small lesions in lung cancer. Combining these findings with other [clinical] data adds another layer of complexity to our diagnostic process.}''

\par \textbf{C3. Providing comprehensive diagnostic results with contextual referencing and comparison is time-consuming.} To deliver well-informed and unbiased diagnostic results, physicians must conduct thorough analyses across various data modalities. This involves using laboratory test results, such as red blood cells, white blood cells, and platelet count, as filters for judgment, and exploring other modalities like image data for further decision-making. However, this cross-modal analysis is time-consuming and adds an extra layer of complexity to the analytical process. As \textbf{E2} mentioned, ``\textit{We have to dive into the nitty-gritty details of diagnostic data, and things get more complicated when we need to combine and analyze [information] from multiple sources.}'' Experienced physicians also acknowledged their reliance on subjective experience for comparison given data with references mentally during diagnosis. However, almost all participants agreed that this comparison, without a clear view, poses challenges, especially for novice physicians' analysis. \textbf{E4} emphasized, ``\textit{As a novice physician, the limited experience makes comparative diagnostic analysis quite a challenge. It's tough to make those subtle comparisons in diagnosis, which comes more naturally to experienced physicians.}''

\par \textbf{C4. Reviewing analyzed data with integrated results and insights leads to additional burdens.} Participants also highlighted the importance of a retrospective view for their learning and analysis during CME. \textbf{E5}, who combines practical teaching with instructing experiences for novice physicians, emphasized, ``\textit{Recalling previous cases helps and draws insights from teaching improves how I diagnose and makes me better overall.}'' However, their current recording methods often separate diagnostic data from corresponding insights and findings, adding extra burdens to their review process. \textbf{E6} further explained, ``\textit{Sorting through scattered information when revisiting cases can be a bit tricky. Extracting meaningful patterns and insights from the collected data becomes a real challenge.}''

\subsection{Design Requirements}
\par Based on insights from the interview analysis, we formulated design requirements tailored to address the identified challenges in analyzing multimodal medical data for physicians. These requirements are structured to align with active, discovery-driven learning processes, guided by the principles of the discovery learning theory~\cite{Bruner1960ThePO}. The structured design requirements are as follows: 1) Initiate an \textbf{overview} level analysis of patient data to foster a broad understanding of concepts built upon prior knowledge (\textbf{R.1-R.2}). 2) Enable users to actively engage in exploration at the \textbf{detail} level for specified cases (\textbf{R.3-R.5}). 3) Facilitate the recording of findings for subsequent review and \textbf{retrospection}, providing scaffolding to consolidate understanding (\textbf{R.6}).

\par {\prefix{\texttt{\textbf{[Overview]}}}}\textbf{R1. Present patient characteristics overview.} Physicians expressed challenges in efficiently filtering and identifying valuable patient cases for retrospective learning. They emphasized the importance of analyzing cohorts with similar conditions or symptoms, crucial for diagnostic decision-making and CME practice enhancement. Furthermore, physicians stressed the significance of studying cohorts with specific demographic characteristics like age and occupation. As articulated by \textbf{E3}, ``\textit{When you're looking at osteoporosis, you'll find that [different] age groups usually have specific common conditions. It helps us understand how factors such as age play a role in the development and progression of diseases like osteoporosis. Besides, the same goes for occupational health. If we check for joint degradation in people with extensive hours of physical labor, it reveals risk factors and ways to prevent issues for certain job-related groups.}''. To mitigate \textbf{C1}, it is imperative to provide an overview to enable physicians to seamlessly view and filter patient cases based on relevant characteristics.

\par {\prefix{\texttt{\textbf{[Overview]}}}}\textbf{R2. Facilitate comparison of multimodal data across patient cases.} Physicians emphasized the importance of being able to compare multimodal data across patient cases. Such comparisons often unveil distinctive insights when different modalities are considered. Specifically, two patient cases might share similarities in indicator data but show significant differences when involving consideration of image data. As shared by \textbf{E5}, ``\textit{For example, we had two patients with blood tests showing a marker called carcinoembryonic antigen (CEA), which can hint at possible cancer risk. But when we dug deeper with MRI scans, we saw some big differences between them. One patient's MRI showed everything [looking] normal - consistent signal intensities and no signs of abnormalities, indicating healthy tissue. But in the other patient's [scan], things were different. We saw irregular signal levels and masses, which could mean there's some abnormal, maybe cancerous tissue there.}'' The multimodal comparison brings out variations, highlighting unique patterns and correlations among different patient cases, thus offering valuable clinical insights.

\par {\prefix{\texttt{\textbf{[Detail]}}}}\textbf{R3. Uncover connections between image and other modalities.} Recognizing the challenges in interpreting image data mentioned in \textbf{C2}, physicians considered it highly beneficial to correlate image-related information with corresponding diagnostic data. For instance, linking relevant areas in an image with corresponding text descriptions could provide valuable insights for novice physicians during the interpretation process. As explained by \textbf{E4}, ``\textit{When we're looking at a spinal MRI, if we can match up the interesting parts in the images with notes that describe specific spinal issues, like disc herniation or spinal stenosis, it really helps us get a much better understanding.''}

\par {\prefix{\texttt{\textbf{[Detail]}}}}\textbf{R4. Provide coherent representation of aligned image data with corresponding information.} Physicians emphasized the need for structured presentation of image-enhanced data, recognizing the benefits of aligning multimodal data for analysis. \textbf{E5} noted, ``\textit{Combining textbook-style explanations with illustrative diagrams provides a clearer picture for analyzing multimodal data in patient cases, especially for guiding new physicians.}''

\par {\prefix{\texttt{\textbf{[Detail]}}}}\textbf{R5. Display relevant contextual information for diagnostic analysis.} Patient cases contain diverse data types beyond images, including structured data and blood test indicators. While displaying essential context is crucial, analyzing these varied datasets comprehensively can be complex and time-consuming. To address \textbf{C3}, different information types should be represented based on their characteristics. This may involve presenting demographic details and patient distribution, visualizing changes in time-series data like medication records, and comparing laboratory results with reference ranges. Though these data may not directly influence diagnostic decisions, presenting them alongside analysis serves as valuable screening references, enhancing efficiency and accuracy.

\par {\prefix{\texttt{\textbf{[Retrospection]}}}}\textbf{R6. Enable recording and reviewing of analysis results.} In response to \textbf{C4}, physicians expressed the need for a more efficient method to document and revisit analyzed insights. For instance, \textbf{E1} highlighted the importance of documenting specific features observed in joint X-rays, like joint space narrowing in osteoarthritis cases, correlating these findings with patient symptoms. This record of insights enhances retrospective learning, crucial in discovery learning for consolidating valuable knowledge. Thus, the ability to record and revisit insightful reflections is essential for improving both the diagnostic process and the physician's knowledge base.

\section{Medillustrator}

\begin{figure}[h]
   \centering
    \vspace{-6mm}
   \includegraphics[width=\columnwidth]{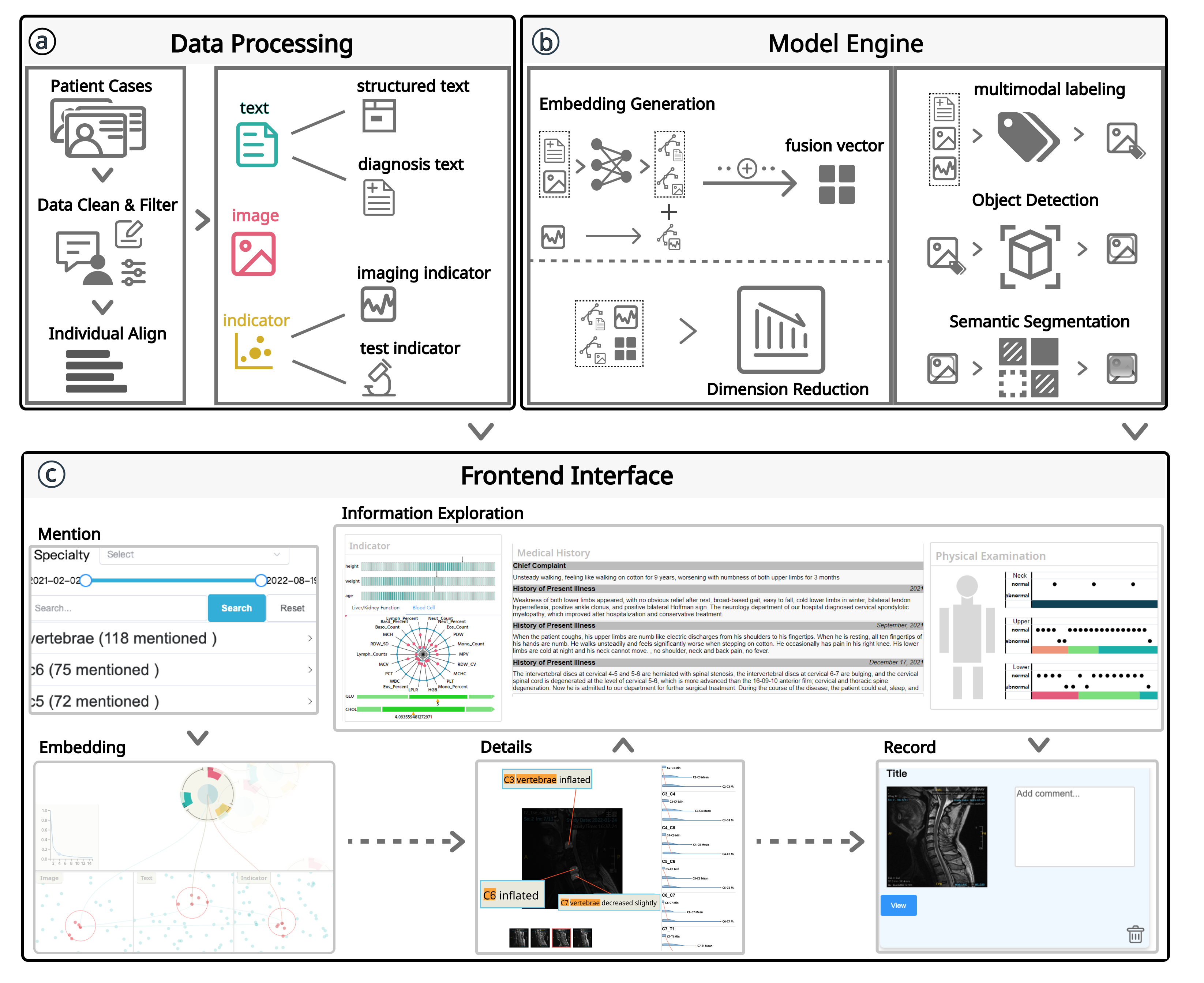}
   \vspace{-6mm}
   \caption{\textit{Medillustrator} is composed of a data process module, a modeling
engine, and a visualization interface designed to align with the discovery-driven learning workflow.}
\vspace{-3mm}
   \label{fig:ApproachOverview}
\end{figure}

\par After gathering discussions and design requirements from the formative study, we developed \textit{Medillustrator}, an interactive visualization system tailored to assist physicians in efficient and effective learning and retrospection using diagnostic data. The system's approach and methods include three core components: (a) a data processing module, (b) a modeling engine, and (c) a visualization interface.


\subsection{Data Processing}
\par To illustrate our approach, we curated a dataset containing diagnosis and treatment records for cervical spine diseases obtained from a reputable local hospital. This dataset consists of data from $720$ patients treated between 2021 and 2022, including MRI images, diagnostic texts, patient demographics, structured data, and more. The data is categorized into three modalities: \textit{image}, \textit{text}, and \textit{indicator}. To protect privacy, we anonymized the data irreversibly. We detailed the processing steps for each modality as follows.

\par \textbf{Image Data Processing:} To model and align the relationship between image data and other modalities, particularly text, a training dataset is essential. We collaborated with physicians to manually annotate MRI images using the \textit{labelme} annotation tool~\cite{russell2008labelme}. Senior physicians identified regions of interest on MRI images during diagnosis, marking them with bounding boxes. They then annotated textual information associated with these areas, specifying the name of the region in the diagnostic text (e.g., specific cervical vertebrae regions and areas of cerebrospinal fluid). This process documented regions of interest alongside corresponding diagnostic text in a \textit{json} format file, facilitating the establishment of connections.

\par \textbf{Text Data Processing:} Each MRI image is paired with a diagnostic text description, including both the "medical description" and "medical diagnosis" sections. In the preprocessing phase, we combined the imaging description and diagnosis, followed by manual cleaning of the text to ensure consistent formatting, aiding alignment modeling.

\par \textbf{Indicator Data Processing:} We began by filtering and categorizing the indicator data into two groups: \textit{imaging indicators} and \textit{laboratory indicators}, based on their characteristics and role in the diagnostic analysis. \textit{Imaging indicators} comprise data from MRI imaging, representing areas like cerebrospinal fluid (CSF) and specific cervical intervertebral disc regions (e.g., C1-C2), with metrics such as maximum, minimum, and average signal intensity values. After consultations with experts, we calculated signal intensity divisions for cervical intervertebral disc regions and the mean of CSF to assess MRI signal intensity. \textit{Laboratory indicators} serve as diagnostic reference points, including measurements related to metabolic processes, organ function (e.g., blood glucose, proteins, enzymes), and immune and health evaluations (e.g., platelets, red blood cells, white blood cell types). These indicators are categorized accordingly. \textit{Imaging indicators} are aligned with other modalities for representation and interactive analysis, while \textit{laboratory indicators} are presented independently for reference.

\subsection{Modeling Engine}
\subsubsection{Multimodal Data Alignment}
\par Effectively aligning diagnostic text descriptions with regions of medical images is crucial for conveying information, especially to novice physicians. To accomplish this, we refined the \textit{grounding dino} model~\cite{Liu2023GroundingDM} within the robust training framework of \textit{mmdetection}. Chosen for its capability as an open-set object detector, the \textit{grounding dino} model excels at recognizing diverse objects based on human inputs like category names or descriptive expressions. Capitalizing on its adaptability, we customized the model to our requirements, enhancing its ability to precisely identify and correlate textual descriptions with corresponding regions in medical images. During training, we curated pairs of image-text data, effectively matching diagnostic descriptions with corresponding image regions. By exposing the model to this annotated dataset, the intricate relationships between textual descriptions and visual features in medical images were learned. The dataset was divided into a training set and a test set with an 8:2 split ratio, resulting in approximately 600 pairs for training and 150 for testing. The training lasted for 50 epochs, achieving a Mean Average Precision (MAP) of 0.53. After training, the model accurately generates bounding box predictions for categorized regions, outlining the areas of interest in the images.

\par To enhance alignment precision and achieve pixel-level correspondence between text and image, we utilized the \textit{segment anything} model~\cite{Kirillov2023SegmentA}. Using the bounding boxes from the \textit{grounding dino} model as inputs, the model performs semantic analysis, generating pixel-level segmentation predictions for relevant regions in the images. This pixel-level alignment improves the visual representation of diagnostic information and aids clearer interpretation, empowering physicians, especially those with less experience, to effectively understand and analyze complex medical data.

\subsubsection{Modality Embedding and Discrepancy Assessment}
\par To provide physicians with a comprehensive overview of patient condition distributions, we integrated information from various modalities, including images, diagnosis text, and \textit{imaging indicators}. For the imaging modality, we utilized the \textit{resnet50} architecture to extract image features, generating 2048-dimensional embeddings. For diagnostic text, we used the \textit{bert-base-uncased} model to obtain 768-dimensional embeddings that capture semantic nuances. For the indicators, we conducted direct dimensionality reduction to synthesize essential quantitative data. These modalities—image, text, and indicators—were then fused by concatenation to form a unified representation, combining rich information from diverse sources. We used \textit{UMAP}~\cite{mcinnes2018umap} to reduce each individual modality and the fused embeddings to two-dimensional coordinates, which equips physicians with enhanced diagnostic capabilities and facilitates informed decision-making.

\par To gain a comprehensive understanding of patient conditions, we explored the interaction between different modalities. Calculating the \textit{k nearest neighbors} (knn) within each patient's modality data coordinates allowed us to understand their profile within each modality's scope. By identifying the intersection of nearest neighbors across modalities, we uncovered areas of agreement and disagreement in patient-level distributions (see \autoref{embedding_view}). This detailed analysis helped identify subtle variations and inconsistencies across modalities, offering valuable insights for clinical decision-making and treatment planning.

\subsection{Visualization Interface}
\par Following the visualization mantra of ``overview first, zoom and filter, then details on demand''~\cite{Shneiderman1996TheEH}, we prioritized usability and effectiveness in our design process. Collaborating with physicians, we refined the \textit{Medillustrator} interface based on formative study findings. The interface comprises five interconnected views: \textit{Mentions View}, \textit{Embedding View}, \textit{Information Exploration View}, \textit{Detail View}, and \textit{Record View}. Aligned with the discovery learning theory~\cite{Bruner1960ThePO}, the interaction workflow is structured into three levels: users begin with a general overview to identify high-value research cases at the \textit{Mentions View} and \textit{Embedding View} (\textbf{R1-R2}). They then delve into comprehensive information exploration about specific patient case patterns at the \textit{Information Exploration View} and \textit{Detail View} (\textbf{R3-R5}). Finally, users record and review analyzed insights for organization and retrospection at the \textit{Record View} (\textbf{R6}).

\subsubsection{Mentions View}
\par The \textit{Mentions View} (\autoref{fig:Teaser}\transparentcircled{a}) offers a comprehensive textual overview of patient data. Users can select their specialty and specify time periods using an intuitive range slider. A search bar allows filtering of patient cases by specific diseases or symptoms. When the search bar is empty, keywords from patient cases are displayed, sorted by frequency, with the most mentioned keyword at the top. Clicking on a keyword reveals related diagnosis texts. Entering specific information in the search bar updates the rows below to show relevant diagnosis texts. Clicking the search icon \raisebox{-0.4ex}{\includegraphics[height=2.2ex]{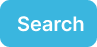}} filters patient cases to only display those containing the entered text in their diagnostic text. This method significantly reduces the number of cases requiring review, enhancing analysis efficiency (\textbf{R1}).

\subsubsection{Embedding View}
\label{embedding_view}
\par
\begin{figure}[h]
    \centering
     \vspace{-3mm}
    \includegraphics[width=\columnwidth]{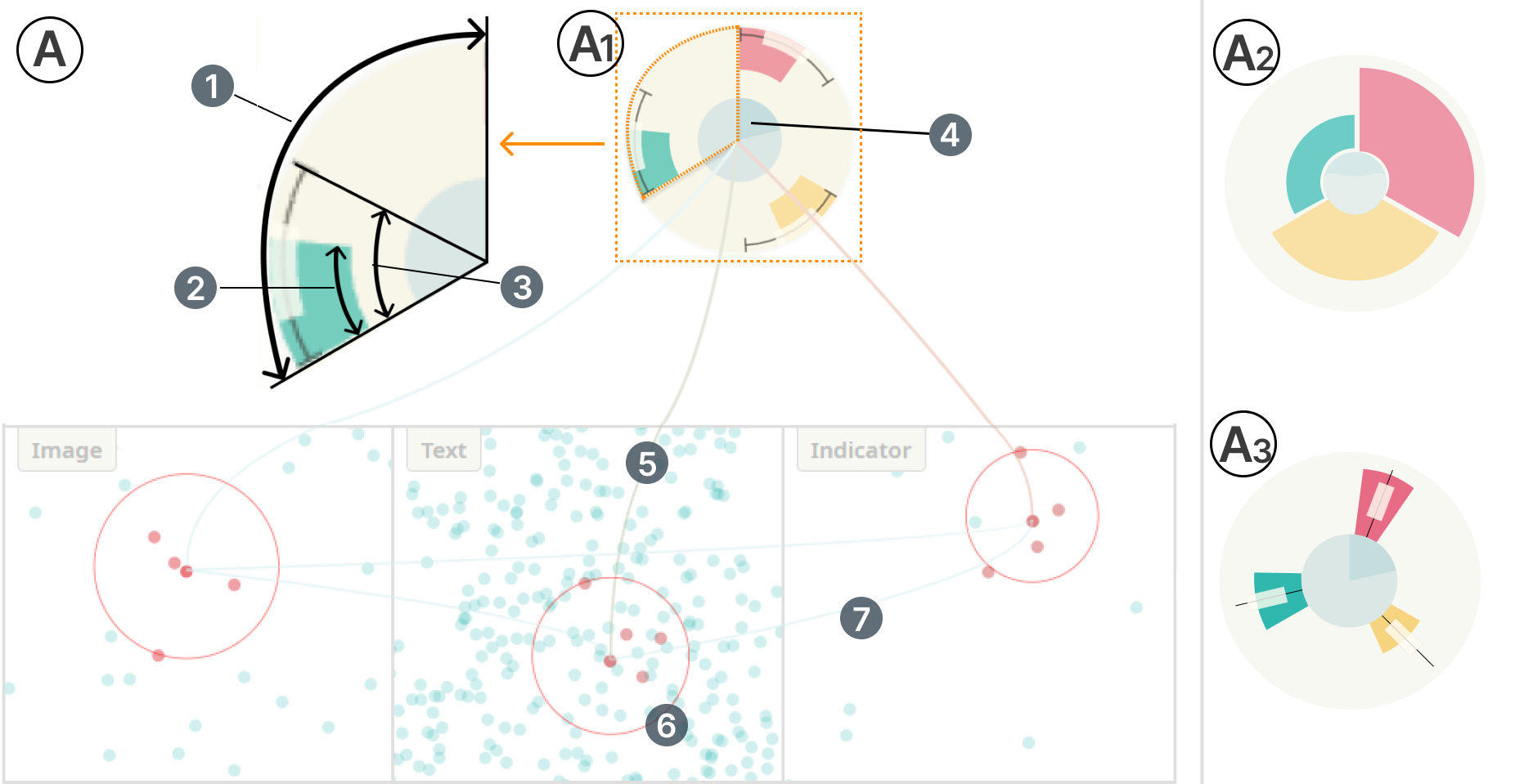}
    \vspace{-6mm}
    \caption{\protect\transparentcircled{A} Glyph design in fusion modal and connections across fusion modal and unimodal. \protect\specialcircled{A}{1} Current glyph design. \protect\specialcircled{A}{2}-\protect\specialcircled{A}{3} Alternatives based on the rose chart and box plot.}
    \label{fig:Embedding}
    \vspace{-3mm}
\end{figure}

 The \textit{Embedding View} (\autoref{fig:Teaser}\transparentcircled{b}) illustrates the distribution of patient cases across modalities' features (\textbf{R2}). Each patient is depicted as a node, showcasing their distribution based on image, text, and indicator features. Additionally, a custom glyph represents the similarity between pairs of these three modalities and their distribution across patients in the fusion modal (\autoref{fig:Embedding}\specialcircled{A}{1}). Selecting a patient by clicking on the glyphs or lassoing for a patient group in the fusion modality connects each glyph to corresponding nodes in other modalities (\autoref{fig:Embedding}\darkcircled{5}). This design allows users to observe patient distribution within the same modality and track patient groups across different modal embeddings. Clicking on a patient's glyph reveals the range and corresponding nodes of the $k$ nearest neighbors (knn) in the three unimodal embedding views below. The \textit{knn} nodes are marked in red within this area, with a circle centered on the current patient's node extending to the farthest \textit{knn} node based on distance (\autoref{fig:Embedding}\darkcircled{6}). Additionally, \textit{knn} nodes corresponding to the same patients across different modalities are interconnected (\autoref{fig:Embedding}\darkcircled{7}). This presentation effectively captures variance information about the surrounding crowds of the same patient across various modalities, facilitating the identification of significant feature differences at different modal levels.

\par To visually represent variations in patient cases with notable differences across modalities, we devised a glyph (\autoref{fig:Embedding}\specialcircled{A}{1}) for each patient case in the fusion modality, depicting the consistency of patient distribution across modalities. The outer ring of the glyph is divided into three equal segments, each occupying 120 degrees (\autoref{fig:Embedding}\darkcircled{1}). These segments denote the similarity among the surrounding crowd between image-text, image-indicator, and text-sentence pairs, respectively. We computed the \textit{knn} set of the current glyph's corresponding patient in the three modalities and utilized the \textit{Jaccard similarity}\cite{Jaccard1912THEDO} calculation method to capture the similarity of the surrounding crowd between these modal pairs. For each segment, we employed a donut chart, where the size of the angle represents the magnitude of the \textit{Jaccard similarity} value under the comparison of the current two modalities (\autoref{fig:Embedding}\darkcircled{2}). Drawing inspiration from existing designs\cite{Ying2021GlyphCreatorTE}, we introduced a refined box plot within each glyph (\autoref{fig:Embedding}\darkcircled{3}), illustrating the distribution of similarity values between the two current modalities among all patients in each segment. This aids in assessing the distribution of modal difference values for the selected patient within the overall patient cohort. In the inner circle area of the glyph, we comprehensively calculated the \textit{Jaccard similarity} of the surrounding patients set across the three modalities ($J(A, B, C) = \frac{|A \cap B \cap C|}{|A \cup B \cup C|}$) and employed a pie chart-based design to depict the overall similarity status of the current glyph's corresponding patient with the surrounding patients in each modality, represented by the size of the fan-shaped angle (\autoref{fig:Embedding}\darkcircled{4}).

\par \textbf{Design Alternatives.} During our design iterations, we explored three glyph designs for the fusion modal. For the first alternative (\autoref{fig:Embedding}\specialcircled{A}{2}), we adapted the rose chart to represent modal similarity. However, we encountered
issues with sectors occupying disproportionate spaces and being closely
connected, hindering user comprehension. For the second alternative (\autoref{fig:Embedding}\specialcircled{A}{3}), we adjusted the angle each pie chart area occupied to better highlight differences in information. Additionally, we incorporated box plots within each sector to depict the distribution across patients. However, user feedback revealed this design could not effectively convey differences in similarity values. Our final design (\autoref{fig:Embedding}\specialcircled{A}{1}) utilized a donut chart, with angle size indicating similarity values, providing clearer representation. Moreover, we transformed the box plot into an arc curve, aligned with the donut chart. The arc’s length indicates similarity distribution across the patient population, with vertical lines denoting minimum and maximum values, and a central box highlighting quartiles. After evaluating all options, we settled on the third design.


\subsubsection{Information Exploration View}
\par We segmented the \textit{Information Exploration View} into three sections, each displaying the \textit{Indicator}, \textit{Medical History}, and \textit{Physical Examination} of the selected patient case. We also simultaneously provided information about the normal reference ranges and the distribution across the general patient population for these indicators (\textbf{R3, R5}).

\par \textbf{Indicator.} In the \textit{indicator} subview (\autoref{fig:Teaser}\specialcircled{C}{1}), we presented patients' \textit{demographic information} and \textit{laboratory indicator}. After consulting domain experts, we selected \textit{height}, \textit{weight}, and \textit{age} as the displayed \textit{demographic information}, ensuring relevance for disease analysis while safeguarding privacy. Using a stripe plot design, color intensity reflects the count of patients corresponding to different indicator values, with a \raisebox{-0.4ex}{\includegraphics[height=2.2ex]{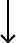}} highlighting the current patient's value.

\par We categorized \textit{laboratory indicators} into: \textit{Renal Function (RF) and Liver Function (LF) Tests} and \textit{Complete Blood Count (CBC)}. Below the \textit{demographic information}, a radar chart displays detailed subcategories and values of \textit{laboratory indicators} for each category. Users can select the specific categories through the menu above the radar chart. The radar chart's outer and inner circles delineate the normal reference range, while red points illustrate the current patient's indicator values. 
This design effectively communicates the overall status of indicator values within each major category while also addressing the need to scrutinize individual indicators of interest.
\par In response to expert feedback emphasizing the importance of certain indicators for disease screening, we introduced stacked bar charts beneath the radar chart. Dark-colored stripes represent the normal reference range, with upward arrows \raisebox{-0.4ex}{\includegraphics[height=2.2ex]{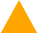}} indicating the current patient's values. 
Following experts' advice, we chose \textit{blood sugar} and \textit{cholesterol} as fixed indicators, enabling users to add new indicators for further analysis by clicking on corresponding nodes \raisebox{-0.4ex}{\includegraphics[height=2.2ex]{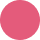}} in the radar chart.

\par \textbf{Design Alternatives.} During the design iteration, we explored various alternatives for presenting \textit{laboratory indicators}. Initially, we tested box plots, beeswarm plots, and violin plots to show current patient values alongside population distribution. However, users prioritize assessing if indicators fall within the normal range for disease analysis, rather than comparing distributions across groups. In the second alternative, we used parallel coordinate axes with normal ranges marked on each axis, but found cluttered lines due to numerous indicators. Our final design employed radar charts, indicating normal ranges in inner and outer circular areas for intuitive assessment and outlier detection. Additionally, we added stacked bar charts based on user feedback to provide detailed insights into specific indicators.

\par \textbf{Medical History.} We organized the \textit{medical history} subview to present the structured text of the current patient. Initially, we highlighted the patient's \textit{Chief Complaint}, offering physicians immediate insights into the primary concern. Subsequently, we presented the patient's medical history information chronologically, starting from the most recent to the oldest, which includes the \textit{History of Present Illness (HPI)}. This chronological sequence aids users in understanding the patient's condition progressively, from the most relevant to the least, thereby streamlining information retrieval efficiency.

\par \textbf{Physical Examination.} In the \textit{Physical Examination} subview (\autoref{fig:Teaser}\transparentcircled{c3}), we presented the patient's physical examination results. After consulting with domain experts, we selected $42$ physical examination indicators covering four areas: \textit{neck}, \textit{upper limbs}, \textit{lower limbs}, and \textit{nervous system}. These indicators can be categorized into three main types: \textit{general condition}, \textit{range of motion}, and \textit{fine movements and reflex testing results}. We found that the qualitative values of these indicators often manifest as binary oppositions, such as normal/abnormal, negative/positive, and presence/absence. To prevent complexity and visual overload in the raw data, we arranged these indicators horizontally in line charts based on different body parts. We utilized "normal" and "abnormal" to represent positive and negative outcomes for each indicator, respectively. Hovering over a specific indicator node on the line chart reveals its name and related information. Additionally, we included a stacked bar chart below the line chart, using differently colored bars to indicate the types of examination. This design allows users to easily differentiate between abnormal and normal indicators within a cluster of various indicators, facilitating the exploration of related information, major categories, and specific examination items.

\subsubsection{Detail View}
\par The \textit{Detail View} (\autoref{fig:Teaser}\transparentcircled{d}) illustrates multimodal medical data alignment and representation based on patient cases selected in the \textit{Embedding View} (\textbf{R3-R4}). Acknowledging the human learning process, especially the forgetting curve~\cite{Ebbinghaus18852013MemoryAC} and cognitive load~\cite{Sweller1998CognitiveAA,Sweller1988CognitiveLD,Paas2003CognitiveLT}, underscores the challenges of flexible knowledge application. To address these challenges, we divided the \textit{Detail View} into two phases: the \textbf{Practice Phase} and the \textbf{Learning Phase}.

\par In the \textbf{Practice Phase}, the raw image of the selected patient case is displayed. Thumbnails of the selected cases appear below, with the currently selected image outlined in \indicatorred{red}. Users can switch between cases by clicking on the respective thumbnails. Additionally, \textit{imaging indicators} are presented on the right side of the image. These indicators include the minimum, maximum, and average signal intensity during MRI imaging process of cervical intervertebral discs (e.g., \textit{C2-C3 = [C2-C3 min, C2-C3 mean, C2-C3 max]}). Using parallel coordinate axes design, the horizontal axis represents the minimum and maximum values of the disc signal indicators among all patients, while a polyline connects the current patient's values across different indicator values within each cervical intervertebral disc region. Density information is incorporated to show the overall distribution on each axis, and indicator data is normalized to reduce biases.

\par Users can visually examine the variation between minimum, average, and maximum values within a specified cervical intervertebral disc region by observing slope change within the group of signal values (e.g., C2-C3). For example, a significant variance between minimum and average values suggests possible disc protrusion. To align \textit{imaging indicators} with the corresponding image areas, we extracted the indicator name as text information to construct connections. Given that our target users are novice physicians with medical backgrounds, we found that their challenge lies in making accurate descriptions and diagnoses rather than matching region names with image areas. Therefore, in the \textit{Practice Phase}, we directly marked the cervical intervertebral disc area on the image with a borderline.

\par Hovering over the relevant area highlights the corresponding indicator, while reducing the opacity of indicators for other parts to minimize disruptions. Users can customize drawing on the image to highlight their area of interest by clicking the \raisebox{-0.4ex}{\includegraphics[height=2.2ex]{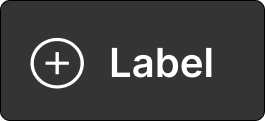}} button, then input labels, diagnostic content, and insights as notes for the selected area. Clicking the \raisebox{-0.4ex}{\includegraphics[height=2.2ex]{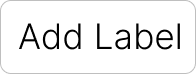}} button marks the selected area on the current image. Users can view added content by hovering over the area or clicking to display it for further editing. After analyzing the current case, users can click the \raisebox{-0.4ex}{\includegraphics[height=2.2ex]{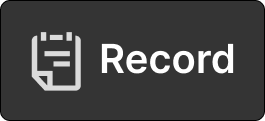}} button to save the patient case and analysis as a card in \textit{Record View} (\autoref{fig:Teaser}\transparentcircled{3}) for subsequent review (\textbf{R6}).



\par Upon completing exploration and analysis in the \textit{Practice Phase}, users can transition to the \textit{Learning Phase} by clicking the \raisebox{-0.2ex}{\includegraphics[height=2.2ex]{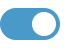}}. In the \textit{Learning Phase}, we established links between relevant areas of the image and corresponding diagnostic text, highlighting these areas by adding a mask to the original image. To facilitate text annotations on the image, we constructed a force-directed model inspired by text layout designs~\cite{Wang2018EdWordleCW, Cui2010ContextPD}. In particular, we proposed the following design principles of the model: \textit{DP1.} Text area should \textit{avoid overlapping} with each other or with the image. \textit{DP2.} Text should be \textit{evenly distributed} around the image, preventing clustering in some areas while leaving others blank. Each diagnostic text and image is treated as a single object in the force simulation and calculation process. Two types of forces are applied to these objects: \textit{textual repulsion forces} and \textit{text-image spring forces}. \textit{Textual repulsion forces} between text area $i$ and $j$, with mass $m_i$ and $m_j$ are proportional to their distance ($F_{i,j} = \frac{m_i \cdot m_j}{r_{i,j}^2}$). This ensures that closer texts experience stronger repulsion, thereby preventing overlaps during position adjustments (\textit{DP1}). Conversely, \textit{text-image spring forces} ($F_i = M \cdot m_i \cdot r_i^2$) between text $i$ and image, with mass $m_i$ and $M$, are inversely proportional to their distance. This maintains a balance, preventing text from straying too far from the image during position adjustments, thus fulfilling \textit{DP2}. The combined effect of these two forces ensures that text areas, subject to balanced forces, are evenly distributed around the image's core. Furthermore, the strength of both forces is proportional to the word count (treated as mass) in the text, considering that texts with more words generally require more display space. To minimize oscillations resulting from the force updates during positional adjustments, we implemented a damping function throughout the iteration process.

\subsubsection{Record View}
\par In the \textit{Record View} (\autoref{fig:Teaser}\transparentcircled{e}), users access previously saved patient cases (\textbf{R6}). Each case is displayed as a card with comprehensive details and analyzed information. Users can input a summary and add descriptive comments on the title area and text area, respectively. Click the \raisebox{-0.2ex}{\includegraphics[height=2.2ex]{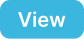}} can switch to previous records and related information in the \textit{Detail} and \textit{Information Exploration View}. They also have the option to export records for further analysis using the export feature (\raisebox{-0.4ex}{\includegraphics[height=2.2ex]{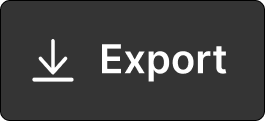}}).

\section{Evaluation}
\par In this section, we performed both a case study and a controlled in-lab between-subject user study to assess the efficacy of the \textit{Medillustrator}.

\subsection{Case Study: Retrospective Learning and Discussion}
\par \textbf{E1}, an intern physician with six months of clinical experience, adopted a meticulous and cautious learning approach while navigating the interaction workflow of \textit{Medillustrator}.

\par
\begin{figure}[h]
    \centering
     \vspace{-3mm}
    \includegraphics[width=\columnwidth]{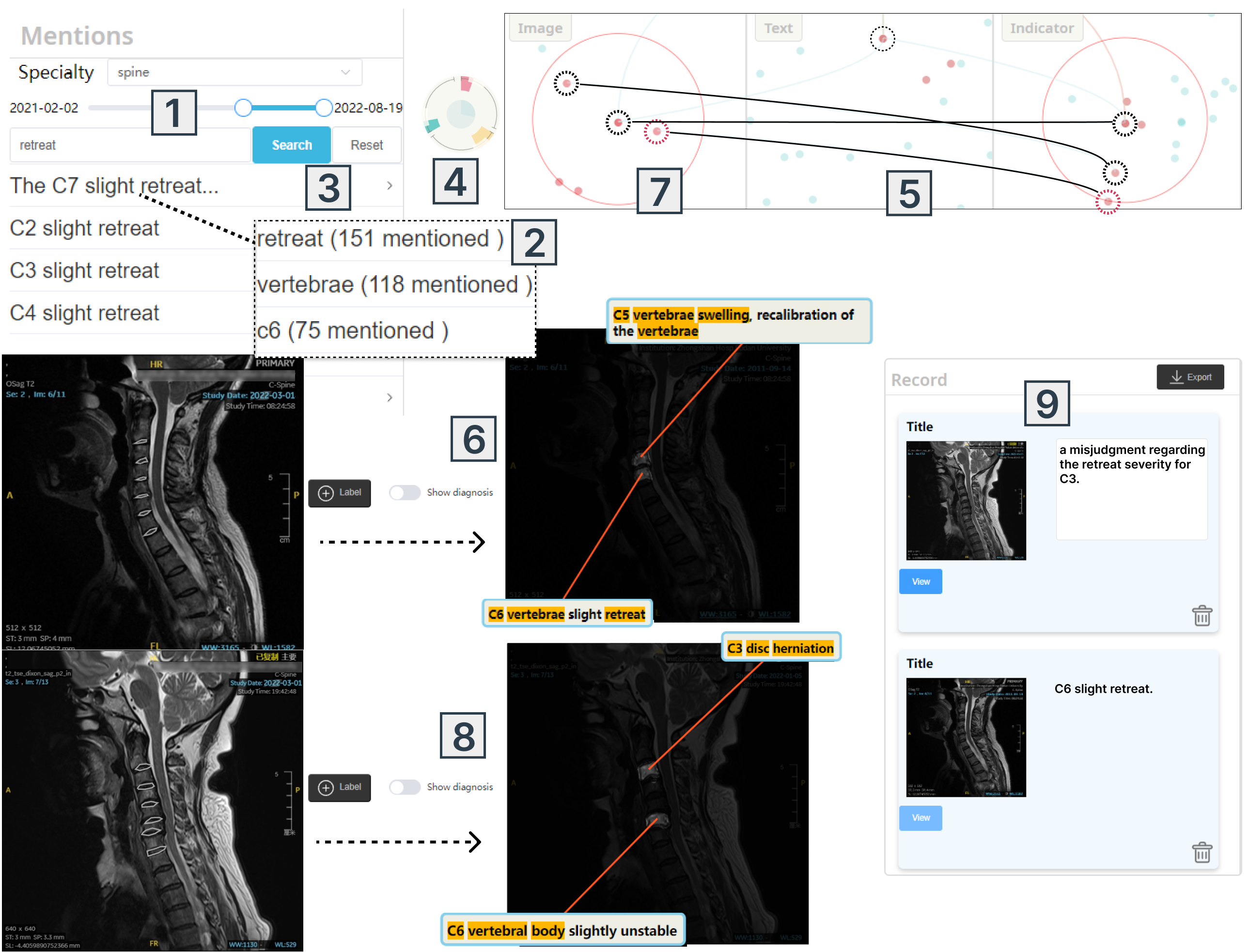}
    \vspace{-6mm}
    \caption{Case Study Part I: Exploring high-value patient cases and addressing interpretation challenges. \protect\casebox{1} Choose the specialty and data time span. \protect\casebox{2} Identify "retreat" as one of the most frequent mentions. \protect\casebox{3} Enter "retreat" in the input box and initiate the search. \protect\casebox{4} Select the relevant glyph in the \textit{Embedding View}. \protect\casebox{5} Notice the simultaneous appearance of three identical k-NNs in both the image and indicator modalities. \protect\casebox{6} Analyze the case in the \textit{Detail View}, annotate findings with \raisebox{-0.4ex}{\includegraphics[height=2.2ex]{figs/add_label.png}}, and review the analysis using \raisebox{-0.2ex}{\includegraphics[height=2.2ex]{figs/slider.png}}. \protect\casebox{7} Navigate to the neighboring case in the image modality and inspect it in the \textit{Detail View}. \protect\casebox{8} Annotate findings with \raisebox{-0.4ex}{\includegraphics[height=2.2ex]{figs/add_label.png}} and review the diagnosis using \raisebox{-0.2ex}{\includegraphics[height=2.2ex]{figs/slider.png}}, discovering an incorrect diagnosis. \protect\casebox{9} Save the results in the \textit{Record View} by clicking \raisebox{-0.4ex}{\includegraphics[height=2.2ex]{figs/record.png}}.}
    \label{fig:case}
    \vspace{-3mm}
\end{figure}

 \textbf{I: Exploring high-value patient cases and acknowledging interpretation challenges.} Beginning by selecting ``spine'' as the specialty of interest and filtering data from the most recent six months (\autoref{fig:case}\protect\casebox{1}), \textbf{E1} scrutinized various mentions presented in the \textit{Mentions View} (\autoref{fig:case}\protect\casebox{2}). Notably, ``retreat'' emerged as one of the top three frequently occurring words, piquing \textbf{E1}'s curiosity. Consequently, \textbf{E1} filtered all diagnostic texts containing this term to investigate associated patient cases (\autoref{fig:case}\protect\casebox{3}).


\par In \textit{Embedding View}, \textbf{E1} noticed one patient glyph conspicuously distant from the others. Upon zooming in, \textbf{E1} observed that while the overall similarity of this case was relatively low, the donut chart in the outer ring depicted a high similarity between the image-indicator, surpassing the three quartiles of patients (\autoref{fig:case}\protect\casebox{4}). Intrigued by this, \textbf{E1} delved deeper by examining the current case's consistency in distribution among neighboring populations in each unimodal. With $k=5$, \textbf{E1} discovered three identical nearest neighbors in both the image and indicator modalities: \textit{c1} (the patient itself), \textit{c2}, and \textit{c3}, but no corresponding matches in the text modality (\autoref{fig:case}\protect\casebox{5}). This discrepancy highlighted a crucial observation: \textit{\textbf{cases with similar images and indicators might ultimately receive disparate diagnostic outcomes}}. This revelation prompted \textbf{E1} to conduct a comprehensive analysis of the current case and its neighbors. Beginning with a closer examination of the current case's image in the \textit{Detail View}, \textbf{E1} identified and marked retreat at the C6 position. Subsequently, clicking on the relevant medical diagnosis revealed confirmation of \textbf{E1}'s initial judgment: ``\textit{C6 vertebrae slight retreat.}'' (\autoref{fig:case}\protect\casebox{6})

\par Turning attention to patient case \textit{c2} in the image modal and clicking for switching (\autoref{fig:case}\protect\casebox{7}), \textbf{E1} meticulously analyzed the raw image and identified a similar protrusion in the \textit{Detail View}. Marking the relevant area during the \textbf{Practice Phase} and referring diagnosis at \textbf{Learning Phase}, \textbf{E1} was surprised to find a different diagnosis for \textit{c2}, with the corresponding parts marked as \textit{disc herniation and slight instability} (\autoref{fig:case}\protect\casebox{8}). Reminded by the diagnostic results, \textbf{E1} recalled the characteristics of disc herniation and instability: while they involve retreat, they denote a more severe degree. This realization led \textbf{E1} to acknowledge a misjudgment regarding the severity of the condition at the initial diagnosis. Reflecting on the analysis, \textbf{E1} recognized the importance of considering ``consistency in some indicators but clear differences in others'', a factor previously overlooked. This insight proved invaluable for accumulating clinical experience and facilitating learning. After completing these analyses, he clicked \raisebox{-0.4ex}{\includegraphics[height=2.2ex]{figs/record.png}} to note insights in \textit{Record View} for future review (\autoref{fig:case}\protect\casebox{9}). In conclusion, \textbf{E1} appreciated how the system facilitated the efficient identification of valuable learning cases, thereby enhancing retrospective learning.

\begin{figure*}
  \centering
  \vspace{-3mm}
  \includegraphics[width=\textwidth]{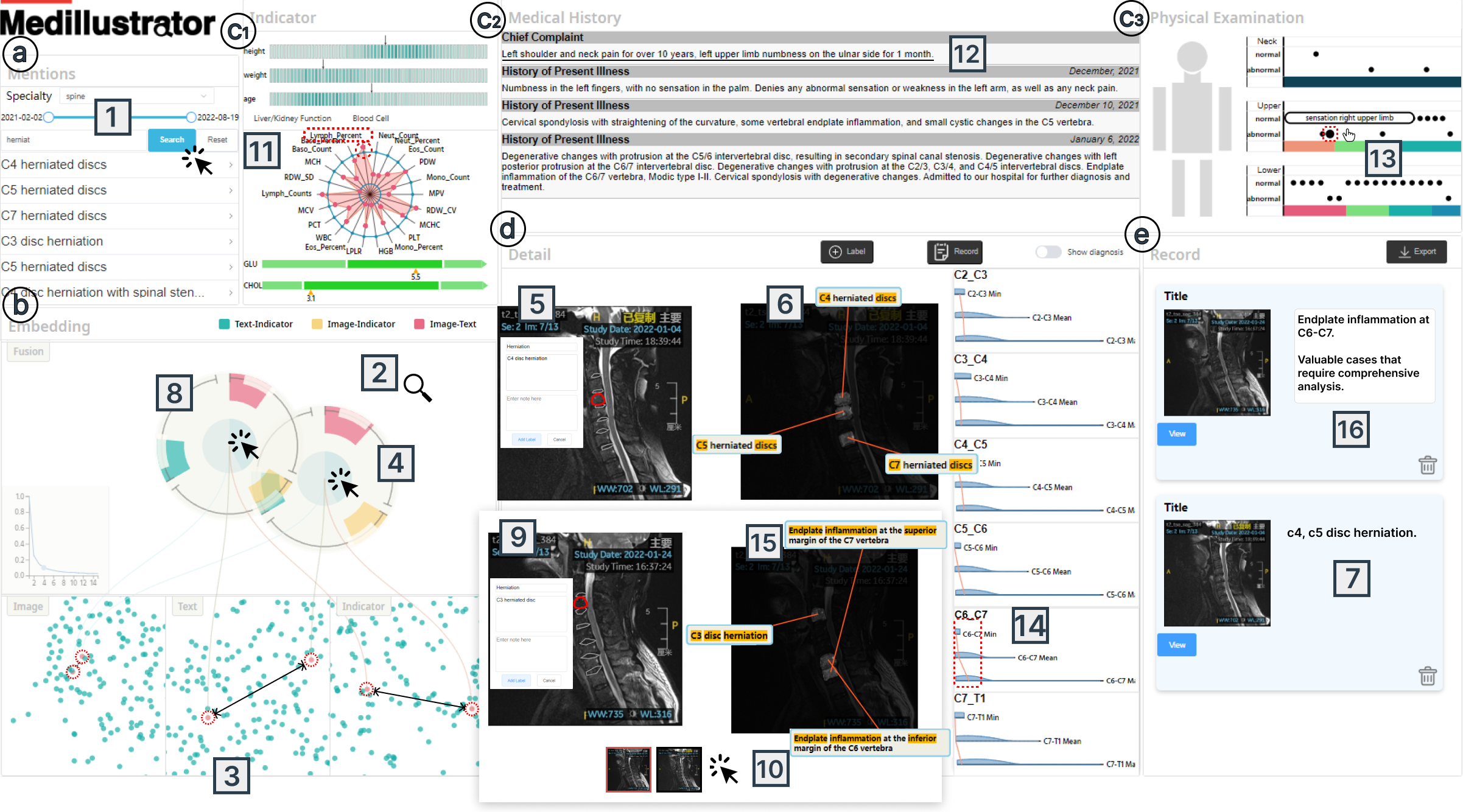}
  \vspace{-6mm}
  \caption{In Case Study Part II, the physicians' interaction workflow comprises: \protect\casebox{1} Filter for cases containing "herniation". \protect\casebox{2} Zoom in to view and discover two close cases (\textit{p1}, \textit{p2}) in \textit{Embedding View}. \protect\casebox{3} Lasso the cases as a group to examine their distribution across modalities, finding the two cases are close in the image modality but distant in the other two modalities. \protect\casebox{4} Click on glyph of case \textit{p1} to examine it in \textit{Detail View}. \protect\casebox{5} Examine and annotate analysis on the image. \protect\casebox{6} Check the aligned diagnosis on the image. \protect\casebox{7} Record the analysis of \textit{p1} in \textit{Record View}. \protect\casebox{8} Click on glyph of case \textit{p2} to examine it in \textit{Detail View}. \protect\casebox{9} Examine the image and annotate partial analysis in \textit{Detail View}. \protect\casebox{10} Switch to check the contextual information between \textit{p1} and \textit{p2} in \textit{Information Exploration View}. \protect\casebox{11} Observe \textit{Indicator} subview and find that the lymphocyte percent of \textit{p2} exceed normal range. \protect\casebox{12} Review \textit{Medical History} subview and find that \textit{p2} has a longer duration of illnesses than \textit{p1}. \protect\casebox{13} Observe that \textit{p2} is examined abnormal in "sensation of right upper limb". \protect\casebox{14} Notice unusual signal values of C6-C7 area in \textit{Detail View}, indicating potential lesions. \protect\casebox{15} Complete the analysis of \textit{p2} and verify its accuracy by checking the aligned diagnosis on the image. \protect\casebox{16} Record the analysis of \textit{p2} in \textit{Record View}.
}
  \label{fig:Teaser}
  \vspace{-2mm}
\end{figure*}

\par \textbf{II: Gaining valuable understandings through the systematic and thorough examination of correlated multimodal patient data.} After completing a previous exploration, \textbf{E1} discussed his learning experience with instructor \textbf{E2} in their weekly meeting. \textbf{E2} recognized the findings and stressed that clinical practice often deviates from textbook scenarios, necessitating thorough and critical analysis. Continuing his learning journey, \textbf{E1} decided to use the system to delve deeper into cases of ``herniated discs'', recalling a previous misdiagnosis. Filtering relevant cases by searching for ``herniated discs'' in the \textit{Mentions View} (\autoref{fig:Teaser}\protect\casebox{1}), \textbf{E1} identified two cases (\textit{p1} and \textit{p2}) exhibiting close proximity in the fusion modal at the \textit{Detail View} (\autoref{fig:Teaser}\protect\casebox{2}). Opting to scrutinize the similarities and differences between these cases, \textbf{E1} zoomed in on the glyphs and selected them as a patient group using lasso. Examining the connected lines between the fusion modal and each unimodal, \textbf{E1} observed that while \textit{p1} and \textit{p2} shared similarities in the image modality, they were notably distinct in the text and indicator modalities (\autoref{fig:Teaser}\protect\casebox{3}). Realizing the significance of studying this pair of cases, \textbf{E1} first clicked on the glyph of \textit{p1} (\autoref{fig:Teaser}\protect\casebox{4}) and examined it in the \textit{Detail View}, identifying varying degrees of disc herniation at C4 and C5. After confirming and reviewing the right-sided corresponding imaging indicator information, \textbf{E1} diagnosed ``c4, c5 disc herniation'' and labeled the relevant area on the image (\autoref{fig:Teaser}\protect\casebox{5}). Clicking to view the diagnosis results further validated his judgment (\autoref{fig:Teaser}\protect\casebox{6}).

\par After adding this cases in the \textit{Record View} (\autoref{fig:Teaser}\protect\casebox{7}), \textbf{E1} clicked to view glyph \textit{p2} (\autoref{fig:Teaser}\protect\casebox{8}). Consistent with the results presented on the \textit{Embedding View}, \textit{p1} and \textit{p2} have similarities in images as they both show the characteristics of disc herniation in certain areas. \textbf{E1} immediately identified that \textit{C3} demonstrated the characteristics of a herniated disc (\autoref{fig:Teaser}\protect\casebox{9}). Learning from the last lesson, \textbf{E1} placed greater emphasis on examining the case comprehensively. Thus, \textbf{E1} toggled between the thumbnails of \textit{p1} and \textit{p2} in the \textit{Record View} (\autoref{fig:Teaser}\protect\casebox{10}) to gather more information before finalizing the diagnostic analysis. \textbf{E1} delved into the contextual reference information within the \textit{Information Exploration View}. In the \textit{Indicator} subview, \textbf{E1} observed that while most of \textit{p1}'s \textit{laboratory indicators} fell within the reference range, the \textit{lymphocyte percentage}, linked to inflammation in \textit{p2}, exceeded the reference values (\autoref{fig:Teaser}\protect\casebox{11}). This prompted \textbf{E1} to consider possible inflammation in \textit{p2}'s body. Moving to the \textit{Medical History} subview, \textbf{E1} noted that both patients complained of finger numbness. However, \textit{p1} experienced symptoms for only a month, whereas \textit{p2} endured them for over ten years, suggesting a longer duration and greater complexity in \textit{p2}'s case (\autoref{fig:Teaser}\protect\casebox{12}). In the \textit{Physical Examination} subview, \textbf{E1} found that \textit{c2} also be examined as abnormal in "Sensation Right Upper Limb" (\autoref{fig:Teaser}\protect\casebox{13}), indicating a much more severity of C2.

\par Upon re-examining \textit{p2}'s corresponding areas in the image at the \textit{Detail View}, \textbf{E1} then noticed that \textit{imaging indicators} contains a relatively low minimum and mean signal value in C6-C7 area, with a significant difference from maximum signal value (\autoref{fig:Teaser}\protect\casebox{14}), hinting at a possible lesion. Inspired by this, \textbf{E1} identified an additional abnormality at C7, resembling end plate inflammation. Including the analysis of ``C7 superior margin end plate inflammation'' in his diagnosis, \textbf{E1} proceeded to validate the diagnoses during the \textbf{Learning Phase} (\autoref{fig:Teaser}\protect\casebox{15}). To his delight, \textbf{E1} confirmed the accuracy of his diagnosis of herniated discs for both \textit{p1} and \textit{p2}. Furthermore, he successfully identified the inflammation of the endplate in \textit{p2}, boosting his confidence. Finally \textbf{E1} added the analysis and insights in the \textit{Record View} (\autoref{fig:Teaser}\protect\casebox{16}). In this analysis, \textbf{E1} highlighted the significance of conducting thorough contextual analyses in diagnosis, confirming the system's value in aiding physicians with efficient analysis. This approach was particularly advantageous for learning and improvement ``\textit{within limited time frames compared to traditional accumulation over extended periods.}''


\subsection{User Study}
\begin{figure*}[h]
    \centering
     \vspace{-3mm}
    \includegraphics[width=\textwidth]{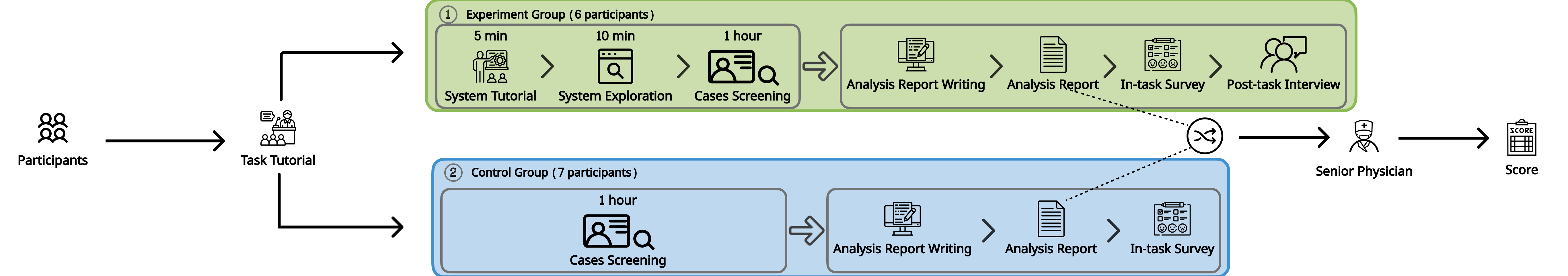}
    \vspace{-6mm}
    \caption{Participants were randomly assigned to two groups: one using \textit{Medillustrator} and the other using a baseline condition for case screening and analysis. The \textit{Medillustrator} group received a tutorial and exploration session to become familiar with the system. Throughout the study, participants completed in-task surveys to evaluate system usability and effectiveness, followed by post-task interviews to gather their perceptions.}
    \label{fig:procedure_user_study}
    \vspace{-3mm}
\end{figure*}

\par \textbf{Baseline System.} We chose the Healthcare Information System (HIS) currently used by physicians for diagnosis as the baseline for comparison. This system enables straightforward reading of structured patient cases, viewing MRI images and reports. We selected it because it mirrors the tool commonly employed by physicians in their daily work for data browsing, analysis, and learning. Its user familiarity and lack of learning curve enable a clear comparison to highlight the enhancements offered by \textit{Medillustrator} over the existing system.

\par \textbf{Participants.} We recruited $13$ participants ($5$ male, $8$ female) via snowball sampling, facilitated by our collaborator \textbf{E3}, a senior physician with over 20 years of experience at a reputable local hospital. To ensure unbiased assessment of user effectiveness and efficiency in using the system for retrospective learning and analysis, we recruited individuals from the same specialty (orthopedics) with similar qualifications. Most participants were interns undergoing standardized training, with one exception—\textbf{P6}, a resident physician with two years of experience. Participants were randomly assigned to two groups: $7$ to the \textit{Medillustrator} condition and $6$ to the baseline system.

\par \textbf{Procedure.} We held discussions with domain experts to identify concerns and challenges regarding CME case analysis. Participants were tasked with screening and analyzing a total of $50$ patient cases curated by two senior physicians, including $10$ cases known for deviations from typical symptom-description patterns found in common ``textbook descriptions''. These $10$ cases served as benchmarks for assessing user selection quality, with participants required to identify the $10$ most valuable cases for retrospective learning. The experiment received IRB approval, and users' feedback was recorded with their consent. Before the experiment, participants underwent a $10$-minute tutorial session led by senior physicians, explaining the tasks and providing examples, with opportunities for questions. Participants in the \textit{Medillustrator} condition received a $5$-minute system tutorial followed by $10$ minutes of exploration. Screening and analysis lasted one hour in both conditions. Participants compiled reports using a given template, outlining their rationale for each selection and insights gained. An in-task questionnaire assessed overall user experience, and participants in the \textit{Medillustrator} condition underwent post-task interviews. The related questions can be found in Appendix. Analysis reports were evaluated by two senior physicians with a multi-observer assessment for consistency~\cite{Shrout1979IntraclassCU}. Participants received $20$ compensation upon completion.

\par \textbf{Measurement.} We utilized a $7$-point Likert scale (from $1$: strongly disagree to $7$: strongly agree) in a post-task questionnaire to evaluate usability and effectiveness. Usability assessment included factors like ease of use, support for case analysis, system-related distractions, user satisfaction, likelihood of recommendation, intent for future use, and perceptions at different analysis stages, inspired by the System Usability Scale (SUS)~\cite{Brooke1996SUSA}. Effectiveness was analyzed based on criteria such as screened case accessibility, efficiency in identifying valuable cases, perceived analysis confidence, recording and reviewing accessibility, and analysis report quality~\cite{hart1988development}. Analysis reports were evaluated for \textit{completeness} (based on the number of identified cases matching the ground truth) and \textit{accuracy} (focused on the precision of the provided rationale for case selection), with detail criterion shown in \autoref{tab:criteria}.

\begin{figure*}[h]
    \centering
    \begin{subfigure}[b]{\textwidth}
        \centering
        \includegraphics[width=\textwidth]{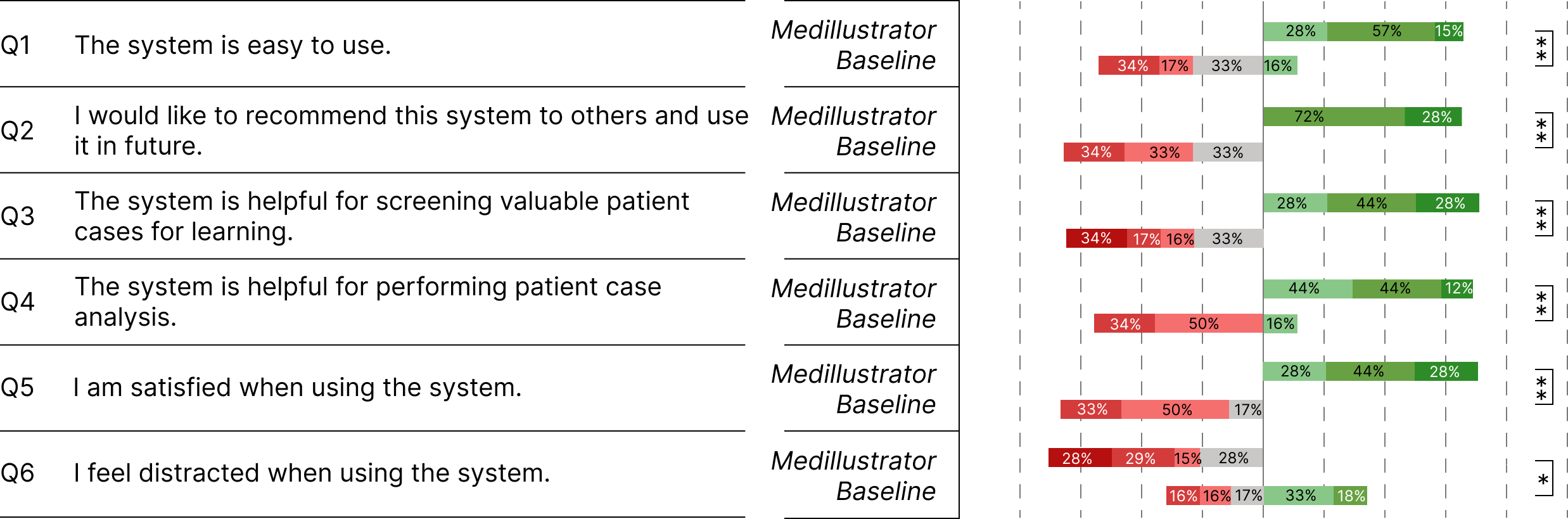}
        \caption{Usability}
        \label{fig:Usability}
    \end{subfigure}
    
    \begin{subfigure}[b]{\textwidth}
        \centering
        \includegraphics[width=\textwidth]{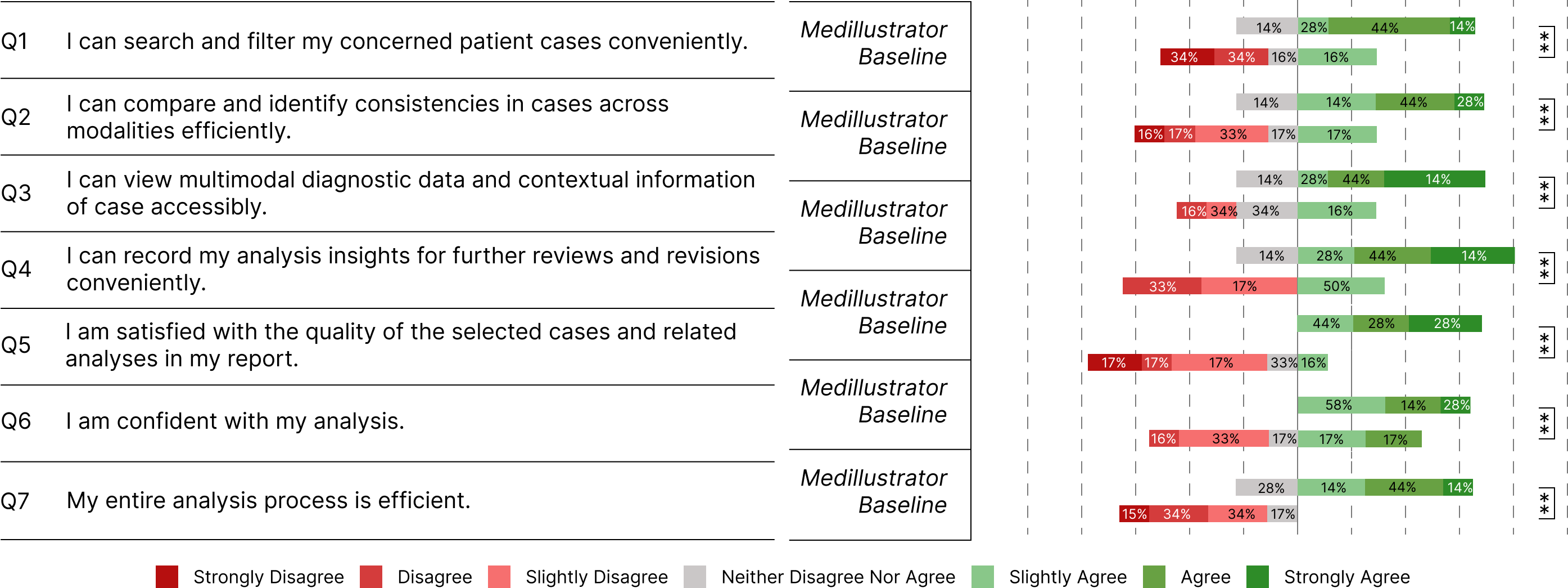}
        \caption{Effectiveness}
        \label{fig:Effectiveness}
    \end{subfigure}
    \vspace{-5mm}
    \caption{In-task survey results from usability and effectiveness aspects in \textit{Medillustrator} and Baseline conditions. (*: p<.05; **: p<.01; ***: p<.001)}
    \label{fig:Results}
    \vspace{-6mm}
\end{figure*}

\begin{table}[h]
\centering
\small
\caption{Comparison of average scores of analysis reports in Completeness and Accuracy of between Control and Experiment Groups.}
\begin{tabular}{
  l
  S[table-format=1.2] 
  S[table-format=1.2] 
  S[table-format=1.1] 
  S[table-format=1.3] 
}
\toprule
 & {Control Group} & {Experiment Group} & {U} & {$p$} \\
\midrule
Completeness & 5.57 & 5.71 & 3   & 0.009 \\
Accuracy     & 3.5  & 4.16 & 3.5 & 0.011 \\
\bottomrule
\end{tabular}
\vspace{-2mm}
\label{tab:Score}
\end{table}

\par \textbf{Results.} We used the \textit{Mann-Whitney U test}~\cite{mann1947test} to compare scores between the \textit{Medillustrator} and baseline systems in both the in-task questionnaire and senior physicians' assessment of analysis reports (\autoref{fig:Results}). Users consistently rated the usability and effectiveness of the \textit{Medillustrator} significantly higher than the baseline. Positive feedback was observed for ease of use (U=1, p < 0.01), likelihood of future recommendations (U=3.5, p < 0.05), helpfulness for screening (U=1, p < 0.01) and analysis of cases (U=3.5, p < 0.01), and satisfaction (U=3.5, p < 0.05). Both systems received relatively low scores for distractions (U=15, p=0.37), indicating no significant differences. Additionally, users acknowledged the \textit{Medillustrator} for its assistance in filtering cases (U=2.5, p < 0.01), identifying inconsistencies (U=2, p < 0.01), and analysis (U=3, p < 0.01). The \textit{Medillustrator} condition showed statistically significant positive feedback in perceived analysis confidence (U=6.5, p < 0.05), efficiency (U=1, p < 0.01), and satisfaction (U=1.5, p < 0.01) in identifying valuable cases. \autoref{tab:Score} shows detailed analysis report ratings, where the \textit{Medillustrator} condition demonstrated superior performance in both completeness (U=3, p < 0.01) and accuracy (U=3.5, p < 0.05). The ICC scores~\cite{Shrout1979IntraclassCU} for each report are above 0.75, indicating good consistency of ratings between reviewers. In post-task interviews, we collected feedback on \textit{Medillustrator}'s features through audio recordings, transcribed, and analyzed them using thematic analysis methods~\cite{Boyatzis1998TransformingQI}. Users appreciated the system's assistance in case screening and analysis with an accessible display. They remarked that ``\textit{Medillustrator is made to fit our' real needs, giving organized and consistent data displays that make analyzing cases easier and more straightforward.}'' Some users (\textbf{P3}, \textbf{P5}, \textbf{P10}) mentioned that grasping the connections and individuals' depiction across different modalities in the \textit{Embedding View} took some getting used to. \textbf{P5} shared, ``\textit{At first, I was a bit confused by the lines and circles in different views, but after a few minutes of explanation and guidance, I started to get the hang of it.}'' However, they also confirmed its effectiveness in case screening, noting, ``\textit{Checking out the glyphs and clicking for details really helps me spot differences between different aspects of cases, making it easier to find those unusual and valuable ones.}'' \textbf{P7} highlighted the usefulness of aligning \textit{imaging indicators} with images in the \textit{Detail View}, saying, ``\textit{When reviewing intervertebral disc signal values, I focus on deviations from the average to the minimum and maximum. The display highlights areas with significant differences, indicating a higher likelihood of protrusion.}'' Users appreciated the option to manually add annotations and make notes during the Practice Phase, stating that ``\textit{Real learning often means getting your hands dirty, not just watching from the sidelines.}'' Additionally, \textbf{P8} suggested adding a ``mistake notebook'' feature for timely recording after correcting errors.

\section{Discussion and Limitation}
\par \textbf{Lessons from Retrospective Case Learning.} In clinical diagnosis, retrospective learning often targets identifying challenging cases that offer insights for learning and research. These cases involve discrepancies between diagnostic data and actual medical conditions, requiring thorough information collection, analysis, and discussion. However, this assumes that physicians have basic knowledge and recognize typical clinical presentations. Alongside difficult cases, retrospective activities may include ``teaching cases'' for medical students early in their clinical diagnosis training. The presence of textbook-style symptoms in these cases and diagnostic data can aid in basic medical education. While \textit{Medillustrator} aids learning from cases, its focus is primarily on analyzing challenging cases with inconsistencies. Future versions can improve case analysis by adding teaching scenarios for interactive learning and offering summarized analyses of similar cases.

\par \textbf{Generalizability and Scalability.} In this work, we primarily focus on cervical spine diagnosis in the orthopedic specialty, a key focus of future work will be exploring the adaptability of this approach to other specialties. By following the same process of image annotation, model fine-tuning, and training, \textit{Medillustrator} could be extended to present aligned displays of images in areas like neurology and pulmonology, further enhancing the learning experience. Regarding visualization capabilities, our system's features extend beyond aiding novice physicians in case analysis learning. For example, the \textit{Embedding View} can identify consistency across modalities, aiding in time-series data analysis, while the \textit{Detail View}'s multimodal data alignment is useful for educational demonstrations. Algorithmically, our methods for aligning images and text semantically facilitate tasks such as medical phrase grounding and visual question answering. To address scalability, hierarchical displays in the \textit{Embedding View} reduce visual clutter, and edge bundling can clarify cross-modal connections.

\par \textbf{Limitations.} Our work has limitations. First, in data acquisition and processing, modality alignment involves manual annotations and large model fine-tuning. Manual annotations could lead to potential risks of limited data diversity due to the restricted specialty of data sources and annotators. One potential solution is to establish a long-term mechanism for interactive annotation and iterative updates. Recording senior physicians' annotations as ongoing data supplements can enhance model training, increase data diversity, and improve novice physicians' learning experience. Additionally, the lack of standardized format in text data can introduce analytical bias into the model due to varying physician descriptive granularity, necessitating a uniform descriptive granularity to mitigate this issue. Second, in our evaluation, while we minimized experimental bias by selecting participants with similar seniority from the same specialty, diverse system perceptions among physicians of different specialties and experience levels are overlooked. Therefore, we plan to conduct continuous studies across various specialties and personnel to evaluate and enhance the system.

\section{Conclusion and Future work}
\par This study identifies challenges and expectations of novice physicians in retrospective learning and introduces \textit{Medillustrator}, an interactive visualization analysis system addressing these needs. \textit{Medillustrator} aids users in identifying valuable cases by illustrating distribution differences across modalities and aligning diagnostic texts with corresponding image regions. Case studies and user evaluations demonstrate \textit{Medillustrator}'s effectiveness in aiding novice physicians to efficiently identify and analyze research cases. Future work involves meeting retrospective learning needs across experience levels and enhancing collaborative analysis.

\begin{acks}
We thank anonymous reviewers for their valuable feedback. This work is supported by grants from the National Natural Science Foundation of China (No. 62372298), Shanghai Frontiers Science Center of Human-centered Artificial Intelligence (ShangHAI), and MoE Key Laboratory of Intelligent Perception and Human-Machine Collaboration (KLIP-HuMaCo).
\end{acks}

\balance
\bibliographystyle{ACM-Reference-Format}
\bibliography{sample-base}


\begin{thebibliography}{71}


\ifx \showCODEN    \undefined \def \showCODEN     #1{\unskip}     \fi
\ifx \showDOI      \undefined \def \showDOI       #1{#1}\fi
\ifx \showISBNx    \undefined \def \showISBNx     #1{\unskip}     \fi
\ifx \showISBNxiii \undefined \def \showISBNxiii  #1{\unskip}     \fi
\ifx \showISSN     \undefined \def \showISSN      #1{\unskip}     \fi
\ifx \showLCCN     \undefined \def \showLCCN      #1{\unskip}     \fi
\ifx \shownote     \undefined \def \shownote      #1{#1}          \fi
\ifx \showarticletitle \undefined \def \showarticletitle #1{#1}   \fi
\ifx \showURL      \undefined \def \showURL       {\relax}        \fi
\providecommand\bibfield[2]{#2}
\providecommand\bibinfo[2]{#2}
\providecommand\natexlab[1]{#1}
\providecommand\showeprint[2][]{arXiv:#2}

\bibitem[317(2020)]%
        {31757278}
 \bibinfo{year}{2020}\natexlab{}.
\newblock \showarticletitle{Active Learning Strategies to Improve Progression
  from Knowledge to Action}.
\newblock  \bibinfo{volume}{46}, \bibinfo{number}{1} (\bibinfo{date}{Feb}
  \bibinfo{year}{2020}), \bibinfo{pages}{1--19}.
\newblock
\showISSN{1558-3163}
\urldef\tempurl%
\url{https://doi.org/10.1016/j.rdc.2019.09.001}
\showDOI{\tempurl}


\bibitem[(1885)(2013)]%
        {Ebbinghaus18852013MemoryAC}
\bibfield{author}{\bibinfo{person}{Hermann~Ebbinghaus (1885)}.}
  \bibinfo{year}{2013}\natexlab{}.
\newblock \showarticletitle{Memory: A Contribution to Experimental Psychology}.
\newblock \bibinfo{journal}{\emph{Annals of Neurosciences}}
  \bibinfo{volume}{20} (\bibinfo{year}{2013}), \bibinfo{pages}{155 -- 156}.
\newblock
\urldef\tempurl%
\url{https://doi.org/10.5214\%2Fans.0972.7531.200408}
\showDOI{\tempurl}


\bibitem[Ahn et~al\mbox{.}(2021)]%
        {Impactofdiagnostic}
\bibfield{author}{\bibinfo{person}{Yura Ahn}, \bibinfo{person}{Gil-Sun Hong},
  \bibinfo{person}{Kye~Jin Park}, \bibinfo{person}{Choong~Wook Lee},
  \bibinfo{person}{Ju~Hee Lee}, {and} \bibinfo{person}{Seon-Ok Kim}.}
  \bibinfo{year}{2021}\natexlab{}.
\newblock \showarticletitle{Impact of diagnostic errors on adverse outcomes:
  learning from emergency department revisits with repeat CT or MRI}.
\newblock \bibinfo{journal}{\emph{Insights into Imaging}} \bibinfo{volume}{12},
  \bibinfo{number}{1} (\bibinfo{year}{2021}), \bibinfo{pages}{160}.
\newblock
\urldef\tempurl%
\url{https://doi.org/10.1186/s13244-021-01108-0}
\showDOI{\tempurl}


\bibitem[Atsina et~al\mbox{.}(2020)]%
        {doi:10.2214/AJR.19.21802}
\bibfield{author}{\bibinfo{person}{Kofi-Buaku Atsina},
  \bibinfo{person}{Laurence Parker}, \bibinfo{person}{Vijay~M. Rao}, {and}
  \bibinfo{person}{David~C. Levin}.} \bibinfo{year}{2020}\natexlab{}.
\newblock \showarticletitle{Advanced Imaging Interpretation by Radiologists and
  Nonradiologist Physicians: A Training Issue}.
\newblock \bibinfo{journal}{\emph{American Journal of Roentgenology}}
  \bibinfo{volume}{214}, \bibinfo{number}{1} (\bibinfo{year}{2020}),
  \bibinfo{pages}{W55--W61}.
\newblock
\urldef\tempurl%
\url{https://doi.org/10.2214/AJR.19.21802}
\showDOI{\tempurl}
\showeprint{https://doi.org/10.2214/AJR.19.21802}
\newblock
\shownote{PMID: 31691611}.


\bibitem[Bannach et~al\mbox{.}(2017)]%
        {Bannach2017VisualAF}
\bibfield{author}{\bibinfo{person}{Andreas Bannach}, \bibinfo{person}{J.
  Bernard}, \bibinfo{person}{Florian Jung}, \bibinfo{person}{J{\"o}rn
  Kohlhammer}, \bibinfo{person}{Thorsten May}, \bibinfo{person}{Kathrin
  Scheckenbach}, {and} \bibinfo{person}{Stefan Wesarg}.}
  \bibinfo{year}{2017}\natexlab{}.
\newblock \showarticletitle{Visual analytics for radiomics: Combining medical
  imaging with patient data for clinical research}.
\newblock \bibinfo{journal}{\emph{2017 IEEE Workshop on Visual Analytics in
  Healthcare (VAHC)}} (\bibinfo{year}{2017}), \bibinfo{pages}{84--91}.
\newblock
\urldef\tempurl%
\url{https://doi.org/10.1109/VAHC.2017.8387545}
\showDOI{\tempurl}


\bibitem[Bordage et~al\mbox{.}(2009)]%
        {Bordage2009ContinuingME}
\bibfield{author}{\bibinfo{person}{Georges Bordage}, \bibinfo{person}{Brian
  Carlin}, {and} \bibinfo{person}{Paul~E. Mazmanian}.}
  \bibinfo{year}{2009}\natexlab{}.
\newblock \showarticletitle{Continuing medical education effect on physician
  knowledge: effectiveness of continuing medical education: American College of
  Chest Physicians Evidence-Based Educational Guidelines.}
\newblock \bibinfo{journal}{\emph{Chest}}  \bibinfo{volume}{135 3 Suppl}
  (\bibinfo{year}{2009}), \bibinfo{pages}{29S--36S}.
\newblock
\urldef\tempurl%
\url{https://doi.org/10.1378/chest.08-2515}
\showDOI{\tempurl}


\bibitem[Boyatzis(1998a)]%
        {boyatzis1998transforming}
\bibfield{author}{\bibinfo{person}{Richard~E Boyatzis}.}
  \bibinfo{year}{1998}\natexlab{a}.
\newblock \bibinfo{booktitle}{\emph{Transforming qualitative information:
  Thematic analysis and code development}}.
\newblock \bibinfo{publisher}{sage}.
\newblock


\bibitem[Boyatzis(1998b)]%
        {Boyatzis1998TransformingQI}
\bibfield{author}{\bibinfo{person}{Richard~E. Boyatzis}.}
  \bibinfo{year}{1998}\natexlab{b}.
\newblock \showarticletitle{Transforming Qualitative Information: Thematic
  Analysis and Code Development}.
\newblock
\urldef\tempurl%
\url{https://api.semanticscholar.org/CorpusID:60526017}
\showURL{%
\tempurl}


\bibitem[Brooke(1996)]%
        {Brooke1996SUSA}
\bibfield{author}{\bibinfo{person}{J.~B. Brooke}.}
  \bibinfo{year}{1996}\natexlab{}.
\newblock \showarticletitle{SUS: A 'Quick and Dirty' Usability Scale}.
\newblock
\urldef\tempurl%
\url{https://api.semanticscholar.org/CorpusID:107686571}
\showURL{%
\tempurl}


\bibitem[Bruner(1960)]%
        {Bruner1960ThePO}
\bibfield{author}{\bibinfo{person}{J{\'e}r{\^o}me~Seymour Bruner}.}
  \bibinfo{year}{1960}\natexlab{}.
\newblock \showarticletitle{The Process of Education}.
\newblock
\urldef\tempurl%
\url{https://api.semanticscholar.org/CorpusID:177285798}
\showURL{%
\tempurl}


\bibitem[Caban and Gotz(2015)]%
        {Caban2015VisualAI}
\bibfield{author}{\bibinfo{person}{Jesus~J. Caban} {and} \bibinfo{person}{David
  Gotz}.} \bibinfo{year}{2015}\natexlab{}.
\newblock \showarticletitle{Visual analytics in healthcare - opportunities and
  research challenges}.
\newblock \bibinfo{journal}{\emph{Journal of the American Medical Informatics
  Association : JAMIA}}  \bibinfo{volume}{22 2} (\bibinfo{year}{2015}),
  \bibinfo{pages}{260--2}.
\newblock
\urldef\tempurl%
\url{https://doi.org/10.1093/jamia/ocv006}
\showDOI{\tempurl}


\bibitem[Chen et~al\mbox{.}(2023)]%
        {Chen2023MedicalPG}
\bibfield{author}{\bibinfo{person}{Zhihao Chen}, \bibinfo{person}{Yangqiaoyu
  Zhou}, \bibinfo{person}{Anh~Tu Tran}, \bibinfo{person}{Junting Zhao},
  \bibinfo{person}{Liang Wan}, \bibinfo{person}{Gideon Su~Kai Ooi},
  \bibinfo{person}{Lionel T.~E. Cheng}, \bibinfo{person}{Choon~Hua Thng},
  \bibinfo{person}{Xinxing Xu}, \bibinfo{person}{Yong Liu}, {and}
  \bibinfo{person}{H. Fu}.} \bibinfo{year}{2023}\natexlab{}.
\newblock \showarticletitle{Medical Phrase Grounding with Region-Phrase Context
  Contrastive Alignment}.
\newblock \bibinfo{journal}{\emph{ArXiv}}  \bibinfo{volume}{abs/2303.07618}
  (\bibinfo{year}{2023}).
\newblock
\urldef\tempurl%
\url{https://doi.org/10.1007/978-3-031-43990-2_35}
\showDOI{\tempurl}


\bibitem[Cheng et~al\mbox{.}(2021)]%
        {cheng2021vbridge}
\bibfield{author}{\bibinfo{person}{Furui Cheng}, \bibinfo{person}{Dongyu Liu},
  \bibinfo{person}{Fan Du}, \bibinfo{person}{Yanna Lin},
  \bibinfo{person}{Alexandra Zytek}, \bibinfo{person}{Haomin Li},
  \bibinfo{person}{Huamin Qu}, {and} \bibinfo{person}{Kalyan Veeramachaneni}.}
  \bibinfo{year}{2021}\natexlab{}.
\newblock \showarticletitle{Vbridge: Connecting the dots between features and
  data to explain healthcare models}.
\newblock \bibinfo{journal}{\emph{IEEE Transactions on Visualization and
  Computer Graphics}} \bibinfo{volume}{28}, \bibinfo{number}{1}
  (\bibinfo{year}{2021}), \bibinfo{pages}{378--388}.
\newblock
\urldef\tempurl%
\url{https://doi.org/10.1109/TVCG.2021.3114836}
\showDOI{\tempurl}


\bibitem[Clarke and Braun(2017)]%
        {clarke2017thematic}
\bibfield{author}{\bibinfo{person}{Victoria Clarke} {and}
  \bibinfo{person}{Virginia Braun}.} \bibinfo{year}{2017}\natexlab{}.
\newblock \showarticletitle{Thematic analysis}.
\newblock \bibinfo{journal}{\emph{The journal of positive psychology}}
  \bibinfo{volume}{12}, \bibinfo{number}{3} (\bibinfo{year}{2017}),
  \bibinfo{pages}{297--298}.
\newblock
\urldef\tempurl%
\url{https://doi.org/doi/10.1037/13620-004}
\showDOI{\tempurl}


\bibitem[Cui et~al\mbox{.}(2022)]%
        {Cui2022DeepMF}
\bibfield{author}{\bibinfo{person}{Can Cui}, \bibinfo{person}{Haichun Yang},
  \bibinfo{person}{Yaohong Wang}, \bibinfo{person}{Shilin Zhao},
  \bibinfo{person}{Zuhayr Asad}, \bibinfo{person}{Lori~A. Coburn},
  \bibinfo{person}{Keith~T. Wilson}, \bibinfo{person}{Bennett~A. Landman},
  {and} \bibinfo{person}{Yuankai Huo}.} \bibinfo{year}{2022}\natexlab{}.
\newblock \showarticletitle{Deep multimodal fusion of image and non-image data
  in disease diagnosis and prognosis: a review}.
\newblock \bibinfo{journal}{\emph{Progress in biomedical engineering (Bristol,
  England)}}  \bibinfo{volume}{5} (\bibinfo{year}{2022}).
\newblock
\urldef\tempurl%
\url{https://doi.org/10.1088/2516-1091/acc2fe}
\showDOI{\tempurl}


\bibitem[Cui et~al\mbox{.}(2010)]%
        {Cui2010ContextPD}
\bibfield{author}{\bibinfo{person}{Weiwei Cui}, \bibinfo{person}{Yingcai Wu},
  \bibinfo{person}{Shixia Liu}, \bibinfo{person}{Furu Wei},
  \bibinfo{person}{Michelle~X. Zhou}, {and} \bibinfo{person}{Huamin Qu}.}
  \bibinfo{year}{2010}\natexlab{}.
\newblock \showarticletitle{Context preserving dynamic word cloud
  visualization}.
\newblock \bibinfo{journal}{\emph{2010 IEEE Pacific Visualization Symposium
  (PacificVis)}} (\bibinfo{year}{2010}), \bibinfo{pages}{121--128}.
\newblock
\urldef\tempurl%
\url{https://doi.org/10.1109/PACIFICVIS.2010.5429600}
\showDOI{\tempurl}


\bibitem[Donkin et~al\mbox{.}(2023)]%
        {2023donkin}
\bibfield{author}{\bibinfo{person}{Rebecca Donkin}, \bibinfo{person}{Heather
  Yule}, {and} \bibinfo{person}{Trina Fyfe}.} \bibinfo{year}{2023}\natexlab{}.
\newblock \showarticletitle{Online case-based learning in medical education: a
  scoping review}.
\newblock \bibinfo{journal}{\emph{BMC Medical Education}} \bibinfo{volume}{23},
  \bibinfo{number}{1} (\bibinfo{year}{2023}), \bibinfo{pages}{564}.
\newblock
\urldef\tempurl%
\url{https://doi.org/10.1186/s12909-023-04520-w}
\showDOI{\tempurl}


\bibitem[Donner-Banzhoff(2018)]%
        {Donner-Banzhoff353}
\bibfield{author}{\bibinfo{person}{Norbert Donner-Banzhoff}.}
  \bibinfo{year}{2018}\natexlab{}.
\newblock \showarticletitle{Solving the Diagnostic Challenge: A
  Patient-Centered Approach}.
\newblock \bibinfo{journal}{\emph{The Annals of Family Medicine}}
  \bibinfo{volume}{16}, \bibinfo{number}{4} (\bibinfo{year}{2018}),
  \bibinfo{pages}{353--358}.
\newblock
\showISSN{1544-1709}
\urldef\tempurl%
\url{https://doi.org/10.1370/afm.2264}
\showDOI{\tempurl}
\showeprint{https://www.annfammed.org/content/16/4/353.full.pdf}


\bibitem[Duanmu et~al\mbox{.}(2020)]%
        {Duanmu2020PredictionOP}
\bibfield{author}{\bibinfo{person}{Hongyi Duanmu},
  \bibinfo{person}{Pauline~Boning Huang}, \bibinfo{person}{Srinidhi Brahmavar},
  \bibinfo{person}{Stephanie Lin}, \bibinfo{person}{Thomas Ren},
  \bibinfo{person}{Jun Kong}, \bibinfo{person}{Fusheng Wang}, {and}
  \bibinfo{person}{Tim~Q. Duong}.} \bibinfo{year}{2020}\natexlab{}.
\newblock \showarticletitle{Prediction of Pathological Complete Response to
  Neoadjuvant Chemotherapy in Breast Cancer Using Deep Learning with
  Integrative Imaging, Molecular and Demographic Data}. In
  \bibinfo{booktitle}{\emph{International Conference on Medical Image Computing
  and Computer-Assisted Intervention}}.
\newblock
\urldef\tempurl%
\url{https://doi.org/10.1007/978-3-030-59713-9_24}
\showDOI{\tempurl}


\bibitem[DynaMed(2024)]%
        {online_videos}
\bibfield{author}{\bibinfo{person}{DynaMed}.} \bibinfo{year}{2024}\natexlab{}.
\newblock \bibinfo{title}{DynaMed}.
\newblock \bibinfo{howpublished}{\url{https://www.dynamed.com/}}.
\newblock


\bibitem[Hart and Staveland(1988)]%
        {hart1988development}
\bibfield{author}{\bibinfo{person}{Sandra~G Hart} {and}
  \bibinfo{person}{Lowell~E Staveland}.} \bibinfo{year}{1988}\natexlab{}.
\newblock \showarticletitle{Development of NASA-TLX (Task Load Index): Results
  of empirical and theoretical research}.
\newblock In \bibinfo{booktitle}{\emph{Advances in psychology}}.
  Vol.~\bibinfo{volume}{52}. \bibinfo{publisher}{Elsevier},
  \bibinfo{pages}{139--183}.
\newblock
\urldef\tempurl%
\url{https://doi.org/10.1016/S0166-4115(08)62386-9}
\showDOI{\tempurl}


\bibitem[Holste et~al\mbox{.}(2021)]%
        {Holste2021EndtoEndLO}
\bibfield{author}{\bibinfo{person}{Greg Holste}, \bibinfo{person}{Savannah~C.
  Partridge}, \bibinfo{person}{Habib Rahbar}, \bibinfo{person}{Debosmita
  Biswas}, \bibinfo{person}{Christoph~I. Lee}, {and} \bibinfo{person}{Adam~M.
  Alessio}.} \bibinfo{year}{2021}\natexlab{}.
\newblock \showarticletitle{End-to-End Learning of Fused Image and Non-Image
  Features for Improved Breast Cancer Classification from MRI}.
\newblock \bibinfo{journal}{\emph{2021 IEEE/CVF International Conference on
  Computer Vision Workshops (ICCVW)}} (\bibinfo{year}{2021}),
  \bibinfo{pages}{3287--3296}.
\newblock
\urldef\tempurl%
\url{https://doi.org/10.1109/ICCVW54120.2021.00368}
\showDOI{\tempurl}


\bibitem[Huang et~al\mbox{.}(2020)]%
        {Huang2020FusionOM}
\bibfield{author}{\bibinfo{person}{Shih-Cheng Huang}, \bibinfo{person}{Anuj
  Pareek}, \bibinfo{person}{Saeed Seyyedi}, \bibinfo{person}{Imon Banerjee},
  {and} \bibinfo{person}{Matthew~P. Lungren}.} \bibinfo{year}{2020}\natexlab{}.
\newblock \showarticletitle{Fusion of medical imaging and electronic health
  records using deep learning: a systematic review and implementation
  guidelines}.
\newblock \bibinfo{journal}{\emph{NPJ Digital Medicine}}  \bibinfo{volume}{3}
  (\bibinfo{year}{2020}).
\newblock
\urldef\tempurl%
\url{https://doi.org/10.1038/s41746-020-00341-z}
\showDOI{\tempurl}


\bibitem[Jaccard(1912)]%
        {Jaccard1912THEDO}
\bibfield{author}{\bibinfo{person}{P. Jaccard}.}
  \bibinfo{year}{1912}\natexlab{}.
\newblock \showarticletitle{THE DISTRIBUTION OF THE FLORA IN THE ALPINE
  ZONE.1}.
\newblock \bibinfo{journal}{\emph{New Phytologist}}  \bibinfo{volume}{11}
  (\bibinfo{year}{1912}), \bibinfo{pages}{37--50}.
\newblock
\urldef\tempurl%
\url{https://doi.org/10.1111/j.1469-8137.1912.tb05611.x}
\showDOI{\tempurl}


\bibitem[Kirillov et~al\mbox{.}(2023)]%
        {Kirillov2023SegmentA}
\bibfield{author}{\bibinfo{person}{Alexander Kirillov}, \bibinfo{person}{Eric
  Mintun}, \bibinfo{person}{Nikhila Ravi}, \bibinfo{person}{Hanzi Mao},
  \bibinfo{person}{Chloe Rolland}, \bibinfo{person}{Laura Gustafson},
  \bibinfo{person}{Tete Xiao}, \bibinfo{person}{Spencer Whitehead},
  \bibinfo{person}{Alexander~C. Berg}, \bibinfo{person}{Wan-Yen Lo},
  \bibinfo{person}{Piotr Doll{\'a}r}, {and} \bibinfo{person}{Ross~B.
  Girshick}.} \bibinfo{year}{2023}\natexlab{}.
\newblock \showarticletitle{Segment Anything}.
\newblock \bibinfo{journal}{\emph{2023 IEEE/CVF International Conference on
  Computer Vision (ICCV)}} (\bibinfo{year}{2023}), \bibinfo{pages}{3992--4003}.
\newblock
\urldef\tempurl%
\url{https://doi.org/10.48550/arXiv.2304.02643}
\showDOI{\tempurl}


\bibitem[Knowles(1975)]%
        {knowles1975self}
\bibfield{author}{\bibinfo{person}{Malcolm~S Knowles}.}
  \bibinfo{year}{1975}\natexlab{}.
\newblock \showarticletitle{Self-directed learning: A guide for learners and
  teachers.}
\newblock  (\bibinfo{year}{1975}).
\newblock


\bibitem[Li et~al\mbox{.}(2023)]%
        {Li2023ACS}
\bibfield{author}{\bibinfo{person}{Yingshu Li}, \bibinfo{person}{Yunyi Liu},
  \bibinfo{person}{Zhanyu Wang}, \bibinfo{person}{Xinyu Liang},
  \bibinfo{person}{Lingqiao Liu}, \bibinfo{person}{Lei Wang},
  \bibinfo{person}{Leyang Cui}, \bibinfo{person}{Zhaopeng Tu},
  \bibinfo{person}{Longyue Wang}, {and} \bibinfo{person}{Luping Zhou}.}
  \bibinfo{year}{2023}\natexlab{}.
\newblock \showarticletitle{A Comprehensive Study of GPT-4V's Multimodal
  Capabilities in Medical Imaging}.
\newblock \bibinfo{journal}{\emph{ArXiv}}  \bibinfo{volume}{abs/2310.20381}
  (\bibinfo{year}{2023}).
\newblock
\urldef\tempurl%
\url{https://doi.org/10.1101/2023.11.03.23298067}
\showDOI{\tempurl}


\bibitem[Liang et~al\mbox{.}(2022)]%
        {Liang2022MultiVizTV}
\bibfield{author}{\bibinfo{person}{Paul~Pu Liang}, \bibinfo{person}{Yiwei Lyu},
  \bibinfo{person}{Gunjan Chhablani}, \bibinfo{person}{Nihal Jain},
  \bibinfo{person}{Zihao Deng}, \bibinfo{person}{Xingbo Wang},
  \bibinfo{person}{Louis-Philippe Morency}, {and} \bibinfo{person}{Ruslan
  Salakhutdinov}.} \bibinfo{year}{2022}\natexlab{}.
\newblock \showarticletitle{MultiViz: Towards Visualizing and Understanding
  Multimodal Models}. In \bibinfo{booktitle}{\emph{International Conference on
  Learning Representations}}.
\newblock
\urldef\tempurl%
\url{https://doi.org/10.48550/arXiv.2207.00056}
\showDOI{\tempurl}


\bibitem[Liu et~al\mbox{.}(2023)]%
        {Liu2023GroundingDM}
\bibfield{author}{\bibinfo{person}{Shilong Liu}, \bibinfo{person}{Zhaoyang
  Zeng}, \bibinfo{person}{Tianhe Ren}, \bibinfo{person}{Feng Li},
  \bibinfo{person}{Hao Zhang}, \bibinfo{person}{Jie Yang},
  \bibinfo{person}{Chun yue Li}, \bibinfo{person}{Jianwei Yang},
  \bibinfo{person}{Hang Su}, \bibinfo{person}{Jun-Juan Zhu}, {and}
  \bibinfo{person}{Lei Zhang}.} \bibinfo{year}{2023}\natexlab{}.
\newblock \showarticletitle{Grounding DINO: Marrying DINO with Grounded
  Pre-Training for Open-Set Object Detection}.
\newblock \bibinfo{journal}{\emph{ArXiv}}  \bibinfo{volume}{abs/2303.05499}
  (\bibinfo{year}{2023}).
\newblock
\urldef\tempurl%
\url{https://doi.org/10.48550/arXiv.2303.05499}
\showDOI{\tempurl}


\bibitem[Liu and Sullivan(2021)]%
        {liu2021story}
\bibfield{author}{\bibinfo{person}{Tzu-Hung Liu} {and} \bibinfo{person}{Amy~M
  Sullivan}.} \bibinfo{year}{2021}\natexlab{}.
\newblock \showarticletitle{A story half told: a qualitative study of medical
  students’ self-directed learning in the clinical setting}.
\newblock \bibinfo{journal}{\emph{BMC Medical Education}} \bibinfo{volume}{21},
  \bibinfo{number}{1} (\bibinfo{year}{2021}), \bibinfo{pages}{1--11}.
\newblock
\urldef\tempurl%
\url{https://doi.org/10.1186/s12909-021-02913-3}
\showDOI{\tempurl}


\bibitem[Lu et~al\mbox{.}(2020)]%
        {Lu2020AIbasedPP}
\bibfield{author}{\bibinfo{person}{Ming~Y. Lu}, \bibinfo{person}{Tiffany~Y.
  Chen}, \bibinfo{person}{Drew F.~K. Williamson}, \bibinfo{person}{Melissa
  Zhao}, \bibinfo{person}{Maha Shady}, \bibinfo{person}{Jana Lipkov{\'a}},
  {and} \bibinfo{person}{Faisal Mahmood}.} \bibinfo{year}{2020}\natexlab{}.
\newblock \showarticletitle{AI-based pathology predicts origins for cancers of
  unknown primary}.
\newblock \bibinfo{journal}{\emph{Nature}}  \bibinfo{volume}{594}
  (\bibinfo{year}{2020}), \bibinfo{pages}{106 -- 110}.
\newblock
\urldef\tempurl%
\url{https://doi.org/10.1038/s41586-021-03512-4}
\showDOI{\tempurl}


\bibitem[Mann and Whitney(1947)]%
        {mann1947test}
\bibfield{author}{\bibinfo{person}{Henry~B Mann} {and}
  \bibinfo{person}{Donald~R Whitney}.} \bibinfo{year}{1947}\natexlab{}.
\newblock \showarticletitle{On a test of whether one of two random variables is
  stochastically larger than the other}.
\newblock \bibinfo{journal}{\emph{The annals of mathematical statistics}}
  (\bibinfo{year}{1947}), \bibinfo{pages}{50--60}.
\newblock
\urldef\tempurl%
\url{https://doi.org/stable/2236101}
\showDOI{\tempurl}


\bibitem[Marinopoulos et~al\mbox{.}(2007)]%
        {Marinopoulos2007EffectivenessOC}
\bibfield{author}{\bibinfo{person}{Spyridon Marinopoulos},
  \bibinfo{person}{Todd Dorman}, \bibinfo{person}{Neda Ratanawongsa},
  \bibinfo{person}{Lisa~M. Wilson}, \bibinfo{person}{Bimal~H. Ashar},
  \bibinfo{person}{Jeffrey~L. Magaziner}, \bibinfo{person}{Redonda~G. Miller},
  \bibinfo{person}{Patricia~A. Thomas}, \bibinfo{person}{Gregory Prokopowicz},
  \bibinfo{person}{Rehan Qayyum}, {and} \bibinfo{person}{Eric~B. Bass}.}
  \bibinfo{year}{2007}\natexlab{}.
\newblock \showarticletitle{Effectiveness of continuing medical education.}
\newblock \bibinfo{journal}{\emph{Evidence report/technology assessment}}
  \bibinfo{volume}{149} (\bibinfo{year}{2007}), \bibinfo{pages}{1--69}.
\newblock
\urldef\tempurl%
\url{https://api.semanticscholar.org/CorpusID:261068639}
\showURL{%
\tempurl}


\bibitem[McInnes et~al\mbox{.}(2018)]%
        {mcinnes2018umap}
\bibfield{author}{\bibinfo{person}{Leland McInnes}, \bibinfo{person}{John
  Healy}, {and} \bibinfo{person}{James Melville}.}
  \bibinfo{year}{2018}\natexlab{}.
\newblock \showarticletitle{Umap: Uniform manifold approximation and projection
  for dimension reduction}.
\newblock \bibinfo{journal}{\emph{arXiv preprint arXiv:1802.03426}}
  (\bibinfo{year}{2018}).
\newblock


\bibitem[McLean(2016)]%
        {mclean2016case}
\bibfield{author}{\bibinfo{person}{Susan~F McLean}.}
  \bibinfo{year}{2016}\natexlab{}.
\newblock \showarticletitle{Case-based learning and its application in medical
  and health-care fields: a review of worldwide literature}.
\newblock \bibinfo{journal}{\emph{Journal of medical education and curricular
  development}}  \bibinfo{volume}{3} (\bibinfo{year}{2016}),
  \bibinfo{pages}{JMECD--S20377}.
\newblock
\urldef\tempurl%
\url{https://doi.org/10.4137/JMECD.S20377}
\showDOI{\tempurl}


\bibitem[Medscape(2024)]%
        {Medscape}
\bibfield{author}{\bibinfo{person}{Medscape}.} \bibinfo{year}{2024}\natexlab{}.
\newblock \bibinfo{title}{Medscape}.
\newblock \bibinfo{howpublished}{\url{https://www.medscape.org/}}.
\newblock


\bibitem[M{\"o}rth et~al\mbox{.}(2020)]%
        {Mrth2020RadExIV}
\bibfield{author}{\bibinfo{person}{Eric M{\"o}rth}, \bibinfo{person}{Kari~S.
  Wagner-Larsen}, \bibinfo{person}{Erlend Hodneland}, \bibinfo{person}{Camilla
  Krakstad}, \bibinfo{person}{Ingfrid~S. Haldorsen}, \bibinfo{person}{Stefan
  Bruckner}, {and} \bibinfo{person}{Noeska~N. Smit}.}
  \bibinfo{year}{2020}\natexlab{}.
\newblock \showarticletitle{RadEx: Integrated Visual Exploration of
  Multiparametric Studies for Radiomic Tumor Profiling}.
\newblock \bibinfo{journal}{\emph{Computer Graphics Forum}}
  \bibinfo{volume}{39} (\bibinfo{year}{2020}).
\newblock
\urldef\tempurl%
\url{https://doi.org/10.1111/cgf.14172}
\showDOI{\tempurl}


\bibitem[Nguyen et~al\mbox{.}(2021)]%
        {2021nguyen}
\bibfield{author}{\bibinfo{person}{Kevin~A. Nguyen}, \bibinfo{person}{Maura
  Borrego}, \bibinfo{person}{Cynthia~J. Finelli}, \bibinfo{person}{Matt
  DeMonbrun}, \bibinfo{person}{Caroline Crockett}, \bibinfo{person}{Sneha
  Tharayil}, \bibinfo{person}{Prateek Shekhar}, \bibinfo{person}{Cynthia
  Waters}, {and} \bibinfo{person}{Robyn Rosenberg}.}
  \bibinfo{year}{2021}\natexlab{}.
\newblock \showarticletitle{Instructor strategies to aid implementation of
  active learning: a systematic literature review}.
\newblock \bibinfo{journal}{\emph{International Journal of STEM Education}}
  \bibinfo{volume}{8}, \bibinfo{number}{1} (\bibinfo{year}{2021}),
  \bibinfo{pages}{9}.
\newblock
\urldef\tempurl%
\url{https://doi.org/10.1186/s40594-021-00270-7}
\showDOI{\tempurl}


\bibitem[Paas et~al\mbox{.}(2003)]%
        {Paas2003CognitiveLT}
\bibfield{author}{\bibinfo{person}{Fred Paas}, \bibinfo{person}{Alexander
  Renkl}, {and} \bibinfo{person}{John Sweller}.}
  \bibinfo{year}{2003}\natexlab{}.
\newblock \showarticletitle{Cognitive Load Theory and Instructional Design:
  Recent Developments}.
\newblock \bibinfo{journal}{\emph{Educational Psychologist}}
  \bibinfo{volume}{38} (\bibinfo{year}{2003}), \bibinfo{pages}{1 -- 4}.
\newblock
\urldef\tempurl%
\url{https://doi.org/10.1207/S15326985EP3801_1}
\showDOI{\tempurl}


\bibitem[Pattni et~al\mbox{.}(2023)]%
        {10.1371/journal.pone.0288474}
\bibfield{author}{\bibinfo{person}{Chandni Pattni}, \bibinfo{person}{Michael
  Scaffidi}, \bibinfo{person}{Juana Li}, \bibinfo{person}{Shai Genis},
  \bibinfo{person}{Nikko Gimpaya}, \bibinfo{person}{Rishad Khan},
  \bibinfo{person}{Rishi Bansal}, \bibinfo{person}{Nazi Torabi},
  \bibinfo{person}{Catharine~M. Walsh}, {and} \bibinfo{person}{Samir~C.
  Grover}.} \bibinfo{year}{2023}\natexlab{}.
\newblock \showarticletitle{Video-based interventions to improve
  self-assessment accuracy among physicians: A systematic review}.
\newblock \bibinfo{journal}{\emph{PLOS ONE}} \bibinfo{volume}{18},
  \bibinfo{number}{7} (\bibinfo{date}{07} \bibinfo{year}{2023}),
  \bibinfo{pages}{1--15}.
\newblock
\urldef\tempurl%
\url{https://doi.org/10.1371/journal.pone.0288474}
\showDOI{\tempurl}


\bibitem[Qiao et~al\mbox{.}(2014)]%
        {Usingcognitivetheory}
\bibfield{author}{\bibinfo{person}{Yu~Qi Qiao}, \bibinfo{person}{Jun Shen},
  \bibinfo{person}{Xiao Liang}, \bibinfo{person}{Song Ding},
  \bibinfo{person}{Fang~Yuan Chen}, \bibinfo{person}{Li Shao},
  \bibinfo{person}{Qing Zheng}, {and} \bibinfo{person}{Zhi~Hua Ran}.}
  \bibinfo{year}{2014}\natexlab{}.
\newblock \showarticletitle{Using cognitive theory to facilitate medical
  education}.
\newblock \bibinfo{journal}{\emph{BMC Medical Education}} \bibinfo{volume}{14},
  \bibinfo{number}{1} (\bibinfo{year}{2014}), \bibinfo{pages}{79}.
\newblock
\urldef\tempurl%
\url{https://doi.org/10.1186/1472-6920-14-79}
\showDOI{\tempurl}


\bibitem[Qin et~al\mbox{.}(2022)]%
        {Qin2022MedicalIU}
\bibfield{author}{\bibinfo{person}{Ziyuan Qin}, \bibinfo{person}{Huahui Yi},
  \bibinfo{person}{Qicheng Lao}, {and} \bibinfo{person}{Kang Li}.}
  \bibinfo{year}{2022}\natexlab{}.
\newblock \showarticletitle{Medical Image Understanding with Pretrained Vision
  Language Models: A Comprehensive Study}.
\newblock \bibinfo{journal}{\emph{ArXiv}}  \bibinfo{volume}{abs/2209.15517}
  (\bibinfo{year}{2022}).
\newblock
\urldef\tempurl%
\url{https://doi.org/10.48550/arXiv.2209.15517}
\showDOI{\tempurl}


\bibitem[Raidou et~al\mbox{.}(2018)]%
        {Raidou2018BladderRV}
\bibfield{author}{\bibinfo{person}{Renata~Georgia Raidou},
  \bibinfo{person}{Oscar Casares-Magaz}, \bibinfo{person}{Artem Amirkhanov},
  \bibinfo{person}{Vitali Moiseenko}, \bibinfo{person}{Ludvig~Paul Muren},
  \bibinfo{person}{John~P. Einck}, \bibinfo{person}{Anna Vilanova}, {and}
  \bibinfo{person}{Eduard Gr{\"o}ller}.} \bibinfo{year}{2018}\natexlab{}.
\newblock \showarticletitle{Bladder Runner: Visual Analytics for the
  Exploration of RT‐Induced Bladder Toxicity in a Cohort Study}.
\newblock \bibinfo{journal}{\emph{Computer Graphics Forum}}
  \bibinfo{volume}{37} (\bibinfo{year}{2018}).
\newblock
\urldef\tempurl%
\url{https://doi.org/10.1111/cgf.13413}
\showDOI{\tempurl}


\bibitem[Rokach(2010)]%
        {Rokach2010EnsemblebasedC}
\bibfield{author}{\bibinfo{person}{Lior Rokach}.}
  \bibinfo{year}{2010}\natexlab{}.
\newblock \showarticletitle{Ensemble-based classifiers}.
\newblock \bibinfo{journal}{\emph{Artificial Intelligence Review}}
  \bibinfo{volume}{33} (\bibinfo{year}{2010}), \bibinfo{pages}{1--39}.
\newblock
\urldef\tempurl%
\url{https://doi.org/10.1007/s10462-009-9124-7}
\showDOI{\tempurl}


\bibitem[Rosendal et~al\mbox{.}(2013)]%
        {BMCPrimaryCare}
\bibfield{author}{\bibinfo{person}{Marianne Rosendal},
  \bibinfo{person}{Dorte~Ejg Jarbøl}, \bibinfo{person}{Anette~Fischer
  Pedersen}, {and} \bibinfo{person}{Rikke~Sand Andersen}.}
  \bibinfo{year}{2013}\natexlab{}.
\newblock \showarticletitle{Multiple perspectives on symptom interpretation in
  primary care research}.
\newblock \bibinfo{journal}{\emph{BMC Family Practice}} \bibinfo{volume}{14},
  \bibinfo{number}{1} (\bibinfo{year}{2013}), \bibinfo{pages}{167}.
\newblock
\urldef\tempurl%
\url{https://doi.org/10.1186/1471-2296-14-167}
\showDOI{\tempurl}


\bibitem[Russell et~al\mbox{.}(2008)]%
        {russell2008labelme}
\bibfield{author}{\bibinfo{person}{Bryan~C Russell}, \bibinfo{person}{Antonio
  Torralba}, \bibinfo{person}{Kevin~P Murphy}, {and} \bibinfo{person}{William~T
  Freeman}.} \bibinfo{year}{2008}\natexlab{}.
\newblock \showarticletitle{LabelMe: a database and web-based tool for image
  annotation}.
\newblock \bibinfo{journal}{\emph{International journal of computer vision}}
  \bibinfo{volume}{77} (\bibinfo{year}{2008}), \bibinfo{pages}{157--173}.
\newblock
\urldef\tempurl%
\url{https://doi.org/10.1007/s11263-007-0090-8}
\showDOI{\tempurl}


\bibitem[Shahar et~al\mbox{.}(2006)]%
        {Shahar2006DistributedII}
\bibfield{author}{\bibinfo{person}{Yuval Shahar}, \bibinfo{person}{Dina
  Goren-Bar}, \bibinfo{person}{David Boaz}, {and} \bibinfo{person}{Gil Tahan}.}
  \bibinfo{year}{2006}\natexlab{}.
\newblock \showarticletitle{Distributed, intelligent, interactive visualization
  and exploration of time-oriented clinical data and their abstractions}.
\newblock \bibinfo{journal}{\emph{Artificial intelligence in medicine}}
  \bibinfo{volume}{38 2} (\bibinfo{year}{2006}), \bibinfo{pages}{115--35}.
\newblock
\urldef\tempurl%
\url{https://doi.org/10.1016/j.artmed.2005.03.001}
\showDOI{\tempurl}


\bibitem[Shehanaz et~al\mbox{.}(2021)]%
        {Shehanaz2021OptimumWM}
\bibfield{author}{\bibinfo{person}{Shaik Shehanaz}, \bibinfo{person}{Ebenezer
  Daniel}, \bibinfo{person}{Sitaramanjaneya~Reddy Guntur}, {and}
  \bibinfo{person}{Sivaji Satrasupalli}.} \bibinfo{year}{2021}\natexlab{}.
\newblock \showarticletitle{Optimum weighted multimodal medical image fusion
  using particle swarm optimization}.
\newblock \bibinfo{journal}{\emph{Optik}}  \bibinfo{volume}{231}
  (\bibinfo{year}{2021}), \bibinfo{pages}{166413}.
\newblock
\urldef\tempurl%
\url{https://doi.org/10.1016/j.ijleo.2021.166413}
\showDOI{\tempurl}


\bibitem[Shimizu(2023)]%
        {Shimizu2023}
\bibfield{author}{\bibinfo{person}{T. Shimizu}.}
  \bibinfo{year}{2023}\natexlab{}.
\newblock \showarticletitle{Twelve tips for physicians’ mastering expertise
  in diagnostic excellence}.
\newblock \bibinfo{journal}{\emph{MedEdPublish}}  \bibinfo{volume}{13}
  (\bibinfo{year}{2023}), \bibinfo{pages}{21}.
\newblock
\urldef\tempurl%
\url{https://doi.org/10.12688/mep.19618.1}
\showDOI{\tempurl}
\newblock
\shownote{Version 1; Peer review: 2 approved with reservations, 1 not
  approved}.


\bibitem[Shneiderman(1996)]%
        {Shneiderman1996TheEH}
\bibfield{author}{\bibinfo{person}{Ben Shneiderman}.}
  \bibinfo{year}{1996}\natexlab{}.
\newblock \showarticletitle{The eyes have it: a task by data type taxonomy for
  information visualizations}.
\newblock \bibinfo{journal}{\emph{Proceedings 1996 IEEE Symposium on Visual
  Languages}} (\bibinfo{year}{1996}), \bibinfo{pages}{336--343}.
\newblock
\urldef\tempurl%
\url{https://doi.org/10.1016/B978-155860915-0/50046-9}
\showDOI{\tempurl}


\bibitem[Shrout and Fleiss(1979)]%
        {Shrout1979IntraclassCU}
\bibfield{author}{\bibinfo{person}{Patrick~E. Shrout} {and}
  \bibinfo{person}{Joseph~L. Fleiss}.} \bibinfo{year}{1979}\natexlab{}.
\newblock \showarticletitle{Intraclass correlations: uses in assessing rater
  reliability.}
\newblock \bibinfo{journal}{\emph{Psychological bulletin}}  \bibinfo{volume}{86
  2} (\bibinfo{year}{1979}), \bibinfo{pages}{420--8}.
\newblock
\urldef\tempurl%
\url{https://doi.org/doi/10.1037/0033-2909.86.2.420}
\showDOI{\tempurl}


\bibitem[Silberberg et~al\mbox{.}(2021)]%
        {Silberberg2021Melding}
\bibfield{author}{\bibinfo{person}{Mina Silberberg},
  \bibinfo{person}{Lawrence~H. Muhlbaier}, \bibinfo{person}{Elaine
  Hart-Brothers}, \bibinfo{person}{Glenda~M. Small}, \bibinfo{person}{Arwen~E.
  Bunce}, \bibinfo{person}{Rupal Patel}, \bibinfo{person}{Seronda Robinson},
  {and} \bibinfo{person}{Sherman~A. James}.} \bibinfo{year}{2021}\natexlab{}.
\newblock \showarticletitle{Melding Multiple Sources of Knowledge: Using Theory
  and Experiential Knowledge to Design a Community Health Intervention Study}.
\newblock \bibinfo{journal}{\emph{Journal of Participatory Research Methods}}
  \bibinfo{volume}{2}, \bibinfo{number}{3} (\bibinfo{date}{17 11}
  \bibinfo{year}{2021}).
\newblock
\urldef\tempurl%
\url{https://doi.org/10.35844/001c.29013}
\showDOI{\tempurl}


\bibitem[Stephenson et~al\mbox{.}(2020)]%
        {stephenson2020}
\bibfield{author}{\bibinfo{person}{Christopher~R. Stephenson},
  \bibinfo{person}{Sara~L. Bonnes}, \bibinfo{person}{Adam~P. Sawatsky},
  \bibinfo{person}{Lukas~W. Richards}, \bibinfo{person}{Cathy~D. Schleck},
  \bibinfo{person}{Jayawant~N. Mandrekar}, \bibinfo{person}{Thomas~J. Beckman},
  {and} \bibinfo{person}{Christopher~M. Wittich}.}
  \bibinfo{year}{2020}\natexlab{}.
\newblock \showarticletitle{The relationship between learner engagement and
  teaching effectiveness: a novel assessment of student engagement in
  continuing medical education}.
\newblock \bibinfo{journal}{\emph{BMC Medical Education}} \bibinfo{volume}{20},
  \bibinfo{number}{1} (\bibinfo{date}{Nov} \bibinfo{year}{2020}),
  \bibinfo{pages}{403}.
\newblock
\urldef\tempurl%
\url{https://doi.org/10.1186/s12909-020-02331-x}
\showDOI{\tempurl}


\bibitem[Sultanum et~al\mbox{.}(2022)]%
        {sultanum2022chartwalk}
\bibfield{author}{\bibinfo{person}{Nicole Sultanum}, \bibinfo{person}{Farooq
  Naeem}, \bibinfo{person}{Michael Brudno}, {and} \bibinfo{person}{Fanny
  Chevalier}.} \bibinfo{year}{2022}\natexlab{}.
\newblock \showarticletitle{ChartWalk: Navigating large collections of text
  notes in electronic health records for clinical chart review}.
\newblock \bibinfo{journal}{\emph{IEEE Transactions on Visualization and
  Computer Graphics}} \bibinfo{volume}{29}, \bibinfo{number}{1}
  (\bibinfo{year}{2022}), \bibinfo{pages}{1244--1254}.
\newblock
\urldef\tempurl%
\url{https://doi.org/10.1109/TVCG.2022.3209444}
\showDOI{\tempurl}


\bibitem[Sweller(1988)]%
        {Sweller1988CognitiveLD}
\bibfield{author}{\bibinfo{person}{John Sweller}.}
  \bibinfo{year}{1988}\natexlab{}.
\newblock \showarticletitle{Cognitive Load During Problem Solving: Effects on
  Learning}.
\newblock \bibinfo{journal}{\emph{Cogn. Sci.}}  \bibinfo{volume}{12}
  (\bibinfo{year}{1988}), \bibinfo{pages}{257--285}.
\newblock
\urldef\tempurl%
\url{https://doi.org/10.1016/0364-0213(88)90023-7}
\showDOI{\tempurl}


\bibitem[Sweller et~al\mbox{.}(1998)]%
        {Sweller1998CognitiveAA}
\bibfield{author}{\bibinfo{person}{John Sweller}, \bibinfo{person}{Jeroen J.~G.
  van Merrienboer}, {and} \bibinfo{person}{Fred Paas}.}
  \bibinfo{year}{1998}\natexlab{}.
\newblock \showarticletitle{Cognitive Architecture and Instructional Design}.
\newblock \bibinfo{journal}{\emph{Educational Psychology Review}}
  \bibinfo{volume}{10} (\bibinfo{year}{1998}), \bibinfo{pages}{251--296}.
\newblock
\urldef\tempurl%
\url{https://doi.org/10.1023/A:1022193728205}
\showDOI{\tempurl}


\bibitem[Tagawa(2008)]%
        {tagawa2008physician}
\bibfield{author}{\bibinfo{person}{Masami Tagawa}.}
  \bibinfo{year}{2008}\natexlab{}.
\newblock \showarticletitle{Physician self-directed learning and education}.
\newblock \bibinfo{journal}{\emph{The Kaohsiung Journal of Medical Sciences}}
  \bibinfo{volume}{24}, \bibinfo{number}{7} (\bibinfo{year}{2008}),
  \bibinfo{pages}{380--385}.
\newblock
\urldef\tempurl%
\url{https://doi.org/10.1016/S1607-551X(08)70136-0}
\showDOI{\tempurl}


\bibitem[Tang et~al\mbox{.}(2021)]%
        {Tang2021VideoModeratorAR}
\bibfield{author}{\bibinfo{person}{Tan Tang}, \bibinfo{person}{Yanhong Wu},
  \bibinfo{person}{Lingyun Yu}, \bibinfo{person}{Yuhong Li}, {and}
  \bibinfo{person}{Yingcai Wu}.} \bibinfo{year}{2021}\natexlab{}.
\newblock \showarticletitle{VideoModerator: A Risk-aware Framework for
  Multimodal Video Moderation in E-Commerce}.
\newblock \bibinfo{journal}{\emph{IEEE Transactions on Visualization and
  Computer Graphics}}  \bibinfo{volume}{PP} (\bibinfo{year}{2021}),
  \bibinfo{pages}{1--1}.
\newblock
\urldef\tempurl%
\url{https://doi.org/10.1109/TVCG.2021.3114781}
\showDOI{\tempurl}


\bibitem[VisualDx(2024)]%
        {VisualDx}
\bibfield{author}{\bibinfo{person}{VisualDx}.} \bibinfo{year}{2024}\natexlab{}.
\newblock \bibinfo{title}{VisualDx}.
\newblock \bibinfo{howpublished}{\url{https://www.visualdx.com/}}.
\newblock


\bibitem[Wang et~al\mbox{.}(2022)]%
        {Wang2022PyMICAD}
\bibfield{author}{\bibinfo{person}{Guotai Wang}, \bibinfo{person}{Xiangde Luo},
  \bibinfo{person}{Ran Gu}, \bibinfo{person}{Shuojue Yang},
  \bibinfo{person}{Yijie Qu}, \bibinfo{person}{Shuwei Zhai},
  \bibinfo{person}{Qianfei Zhao}, \bibinfo{person}{Kang Li}, {and}
  \bibinfo{person}{Shaoting Zhang}.} \bibinfo{year}{2022}\natexlab{}.
\newblock \showarticletitle{PyMIC: A deep learning toolkit for
  annotation-efficient medical image segmentation}.
\newblock \bibinfo{journal}{\emph{Computer methods and programs in
  biomedicine}}  \bibinfo{volume}{231} (\bibinfo{year}{2022}),
  \bibinfo{pages}{107398}.
\newblock
\urldef\tempurl%
\url{https://doi.org/10.1016/j.cmpb.2023.107398}
\showDOI{\tempurl}


\bibitem[Wang et~al\mbox{.}(2021c)]%
        {Wang2021ModelingUI}
\bibfield{author}{\bibinfo{person}{Hongzhi Wang}, \bibinfo{person}{Vaishnavi
  Subramanian}, {and} \bibinfo{person}{Tanveer~F. Syeda-Mahmood}.}
  \bibinfo{year}{2021}\natexlab{c}.
\newblock \showarticletitle{Modeling Uncertainty in Multi-Modal Fusion for Lung
  Cancer Survival Analysis}.
\newblock \bibinfo{journal}{\emph{2021 IEEE 18th International Symposium on
  Biomedical Imaging (ISBI)}} (\bibinfo{year}{2021}),
  \bibinfo{pages}{1169--1172}.
\newblock
\urldef\tempurl%
\url{https://doi.org/10.1109/ISBI48211.2021.9433823}
\showDOI{\tempurl}


\bibitem[Wang et~al\mbox{.}(2021b)]%
        {Wang2021ThreadStatesSV}
\bibfield{author}{\bibinfo{person}{Qianwen Wang}, \bibinfo{person}{Tali Mazor},
  \bibinfo{person}{Theresa~Anisja Harbig}, \bibinfo{person}{Ethan~G. Cerami},
  {and} \bibinfo{person}{Nils Gehlenborg}.} \bibinfo{year}{2021}\natexlab{b}.
\newblock \showarticletitle{ThreadStates: State-based Visual Analysis of
  Disease Progression}.
\newblock \bibinfo{journal}{\emph{IEEE Transactions on Visualization and
  Computer Graphics}}  \bibinfo{volume}{PP} (\bibinfo{year}{2021}),
  \bibinfo{pages}{1--1}.
\newblock
\urldef\tempurl%
\url{https://doi.org/10.1109/TVCG.2021.3114840}
\showDOI{\tempurl}


\bibitem[Wang et~al\mbox{.}(2020)]%
        {Wang2020AnnotationefficientDL}
\bibfield{author}{\bibinfo{person}{Shanshan Wang}, \bibinfo{person}{Cheng Li},
  \bibinfo{person}{Rongpin Wang}, \bibinfo{person}{Zaiyi Liu},
  \bibinfo{person}{Meiyun Wang}, \bibinfo{person}{Hongna Tan},
  \bibinfo{person}{Yaping Wu}, \bibinfo{person}{Xinfeng Liu},
  \bibinfo{person}{Hui Sun}, \bibinfo{person}{Rui Yang}, \bibinfo{person}{Xin
  Liu}, \bibinfo{person}{Jie Chen}, \bibinfo{person}{Hui-Chong Zhou},
  \bibinfo{person}{Ismail~Ben Ayed}, {and} \bibinfo{person}{Hairong Zheng}.}
  \bibinfo{year}{2020}\natexlab{}.
\newblock \showarticletitle{Annotation-efficient deep learning for automatic
  medical image segmentation}.
\newblock \bibinfo{journal}{\emph{Nature Communications}}  \bibinfo{volume}{12}
  (\bibinfo{year}{2020}).
\newblock
\urldef\tempurl%
\url{https://doi.org/10.1038/s41467-021-26216-9}
\showDOI{\tempurl}


\bibitem[Wang et~al\mbox{.}(2021a)]%
        {Wang2021M2LensVA}
\bibfield{author}{\bibinfo{person}{Xingbo Wang}, \bibinfo{person}{Jianben He},
  \bibinfo{person}{Zhihua Jin}, \bibinfo{person}{Muqiao Yang}, {and}
  \bibinfo{person}{Huamin Qu}.} \bibinfo{year}{2021}\natexlab{a}.
\newblock \showarticletitle{M2Lens: Visualizing and Explaining Multimodal
  Models for Sentiment Analysis}.
\newblock \bibinfo{journal}{\emph{IEEE Transactions on Visualization and
  Computer Graphics}}  \bibinfo{volume}{PP} (\bibinfo{year}{2021}),
  \bibinfo{pages}{1--1}.
\newblock
\urldef\tempurl%
\url{https://doi.org/10.1109/TVCG.2021.3114794}
\showDOI{\tempurl}


\bibitem[Wang et~al\mbox{.}(2018)]%
        {Wang2018EdWordleCW}
\bibfield{author}{\bibinfo{person}{Yunhai Wang}, \bibinfo{person}{Xiaowei Chu},
  \bibinfo{person}{Chen Bao}, \bibinfo{person}{Lifeng Zhu},
  \bibinfo{person}{Oliver Deussen}, \bibinfo{person}{Baoquan Chen}, {and}
  \bibinfo{person}{Michael Sedlmair}.} \bibinfo{year}{2018}\natexlab{}.
\newblock \showarticletitle{EdWordle: Consistency-Preserving Word Cloud
  Editing}.
\newblock \bibinfo{journal}{\emph{IEEE Transactions on Visualization and
  Computer Graphics}}  \bibinfo{volume}{24} (\bibinfo{year}{2018}),
  \bibinfo{pages}{647--656}.
\newblock
\urldef\tempurl%
\url{https://doi.org/10.1109/TVCG.2017.2745859}
\showDOI{\tempurl}


\bibitem[Worum et~al\mbox{.}(2019)]%
        {Bridgingthegap}
\bibfield{author}{\bibinfo{person}{Hilde Worum}, \bibinfo{person}{Daniela
  Lillekroken}, \bibinfo{person}{Birgitte Ahlsen},
  \bibinfo{person}{Kirsti~Skavberg Roaldsen}, {and} \bibinfo{person}{Astrid
  Bergland}.} \bibinfo{year}{2019}\natexlab{}.
\newblock \showarticletitle{Bridging the gap between research-based knowledge
  and clinical practice: a qualitative examination of patients and
  physiotherapists’ views on the Otago exercise Programme}.
\newblock \bibinfo{journal}{\emph{BMC Geriatrics}} \bibinfo{volume}{19},
  \bibinfo{number}{1} (\bibinfo{year}{2019}), \bibinfo{pages}{278}.
\newblock
\urldef\tempurl%
\url{https://doi.org/10.1186/s12877-019-1309-6}
\showDOI{\tempurl}


\bibitem[Wu et~al\mbox{.}(2023)]%
        {Wu2023LiveRetroVA}
\bibfield{author}{\bibinfo{person}{Yuchen Wu}, \bibinfo{person}{Yuansong Xu},
  \bibinfo{person}{Shenghan Gao}, \bibinfo{person}{Xingbo Wang},
  \bibinfo{person}{Wen gang Song}, \bibinfo{person}{Zhiheng Nie},
  \bibinfo{person}{X. Fan}, {and} \bibinfo{person}{Qu Li}.}
  \bibinfo{year}{2023}\natexlab{}.
\newblock \showarticletitle{LiveRetro: Visual Analytics for Strategic
  Retrospect in Livestream E-Commerce}.
\newblock \bibinfo{journal}{\emph{IEEE Transactions on Visualization and
  Computer Graphics}}  \bibinfo{volume}{30} (\bibinfo{year}{2023}),
  \bibinfo{pages}{1117--1127}.
\newblock
\urldef\tempurl%
\url{https://doi.org/10.1109/TVCG.2023.3326911}
\showDOI{\tempurl}


\bibitem[Yang et~al\mbox{.}(2023)]%
        {Yang2023LeveragingHM}
\bibfield{author}{\bibinfo{person}{Ouyang Yang}, \bibinfo{person}{Yuchen Wu},
  \bibinfo{person}{Hengbao Wang}, \bibinfo{person}{Chenyang Zhang},
  \bibinfo{person}{Furui Cheng}, \bibinfo{person}{Chang Jiang},
  \bibinfo{person}{Lixia Jin}, \bibinfo{person}{Yuanwu Cao}, {and}
  \bibinfo{person}{Qu Li}.} \bibinfo{year}{2023}\natexlab{}.
\newblock \showarticletitle{Leveraging Historical Medical Records as a Proxy
  via Multimodal Modeling and Visualization to Enrich Medical Diagnostic
  Learning}.
\newblock \bibinfo{journal}{\emph{IEEE transactions on visualization and
  computer graphics}}  \bibinfo{volume}{PP} (\bibinfo{year}{2023}).
\newblock
\urldef\tempurl%
\url{https://doi.org/10.1109/TVCG.2023.3326929}
\showDOI{\tempurl}


\bibitem[Ying et~al\mbox{.}(2021)]%
        {Ying2021GlyphCreatorTE}
\bibfield{author}{\bibinfo{person}{Lu Ying}, \bibinfo{person}{Tan Tang},
  \bibinfo{person}{Yuzhe Luo}, \bibinfo{person}{Lvkeshen Shen},
  \bibinfo{person}{Xiao Xie}, \bibinfo{person}{Lingyun Yu}, {and}
  \bibinfo{person}{Yingcai Wu}.} \bibinfo{year}{2021}\natexlab{}.
\newblock \showarticletitle{GlyphCreator: Towards Example-based Automatic
  Generation of Circular Glyphs}.
\newblock \bibinfo{journal}{\emph{IEEE Transactions on Visualization and
  Computer Graphics}}  \bibinfo{volume}{PP} (\bibinfo{year}{2021}),
  \bibinfo{pages}{1--1}.
\newblock
\urldef\tempurl%
\url{https://doi.org/10.1109/TVCG.2021.3114877}
\showDOI{\tempurl}


\bibitem[Zeng et~al\mbox{.}(2022)]%
        {Zeng2022GestureLensVA}
\bibfield{author}{\bibinfo{person}{Haipeng Zeng}, \bibinfo{person}{Xingbo
  Wang}, \bibinfo{person}{Yong Wang}, \bibinfo{person}{Aoyu Wu},
  \bibinfo{person}{Ting-Chuen Pong}, {and} \bibinfo{person}{Huamin Qu}.}
  \bibinfo{year}{2022}\natexlab{}.
\newblock \showarticletitle{GestureLens: Visual Analysis of Gestures in
  Presentation Videos}.
\newblock \bibinfo{journal}{\emph{IEEE Transactions on Visualization and
  Computer Graphics}}  \bibinfo{volume}{PP} (\bibinfo{year}{2022}),
  \bibinfo{pages}{1--1}.
\newblock
\urldef\tempurl%
\url{https://doi.org/10.1109/TVCG.2022.3169175}
\showDOI{\tempurl}


\bibitem[Zhu et~al\mbox{.}(2022)]%
        {Zhu2022MedicalLS}
\bibfield{author}{\bibinfo{person}{Xiner Zhu}, \bibinfo{person}{Yichao Wu},
  \bibinfo{person}{Haoji Hu}, \bibinfo{person}{Xianwei Zhuang},
  \bibinfo{person}{Jincao Yao}, \bibinfo{person}{Di Ou}, \bibinfo{person}{Wei
  Li}, \bibinfo{person}{Mei Song}, \bibinfo{person}{Na Feng}, {and}
  \bibinfo{person}{Dong-Guo Xu}.} \bibinfo{year}{2022}\natexlab{}.
\newblock \showarticletitle{Medical lesion segmentation by combining
  multi-modal images with modality weighted UNet.}
\newblock \bibinfo{journal}{\emph{Medical physics}} (\bibinfo{year}{2022}).
\newblock
\urldef\tempurl%
\url{https://doi.org/10.1002/mp.15610}
\showDOI{\tempurl}


\end{thebibliography}

\appendix
\clearpage
\section{Usability and Effectiveness Questions}
\begin{table}[h]
\centering
\begin{threeparttable}
\label{tab:In_task_questions}
\begin{tabularx}{\textwidth}{l X}
\hline\hline
\textbf{Category} & \textbf{Question} \\
\hline
\multirow{6}{*}{Usability} & 1. The system is easy to use. \\
 & 2. I would like to recommend this system to others and use it in the future. \\
 & 3. The system is helpful for screening valuable patient cases for learning. \\
 & 4. The system is helpful for performing patient case analysis.  \\
 & 5. I am satisfied when using the system. \\
 & 6. I feel distracted when using the system. \\
 \hline
\multirow{7}{*}{Effectiveness} & 1. I can search and filter my concerned patient cases conveniently. \\
 & 2. I can compare and identify consistencies in cases across modalities efficiently. \\
 & 3. I can view multimodal diagnostic data and contextual information of case accessibly. \\
 & 4. I can record my analysis insights for further reviews and revisions conveniently. \\
 & 5. I am satisfied with the quality of the selected cases and related analyses in my report. \\
 & 6. I am confident with my analysis. \\
 & 7. My entire analysis process is efficient.\\
\hline\hline
\end{tabularx}
\caption{In-task survey for participants in 7-point Likert scale(1: Strongly Disagree, 7: Strongly Agree).}
\end{threeparttable}
\end{table}

\clearpage
\section{Criterion of Analysis Report}
\begin{table}[h!]
\centering
\begin{tabular}{>{\raggedright\arraybackslash}p{2cm}>{\raggedright\arraybackslash}p{4cm}>{\raggedright\arraybackslash}p{8cm}}
\hline
\textbf{Criterion} & \textbf{Description} & \textbf{Scoring Criteria} \\ \hline

\multirow{5}{*}{\textbf{Completeness}} 
& \multirow{5}{4cm}{Measures the number of identified cases matching the ground truth.} 
& 9-10: Identifies 95-100\% of cases. \\ \cline{3-3} 
& & 7-8: Identifies 85-94\% of cases. \\ \cline{3-3} 
& & 5-6: Identifies 70-84\% of cases. \\ \cline{3-3} 
& & 3-4: Identifies 50-69\% of cases. \\ \cline{3-3} 
& & 1-2: Identifies less than 50\% of cases. \\ \hline

\multirow{5}{*}{\textbf{Accuracy}} 
& \multirow{5}{4cm}{Assesses the precision of the rationale provided for each selected case.} 
& 9-10: Rationale is precise, relevant, and aligns perfectly with each case. \\ \cline{3-3} 
& & 7-8: Rationale is mostly precise with minor inconsistencies. \\ \cline{3-3} 
& & 5-6: Rationale is somewhat precise but lacks clarity or relevance in some cases \\ \cline{3-3} 
& & 3-4: Rationale is generally imprecise with limited relevance. \\ \cline{3-3} 
& & 1-2: Rationale is inaccurate or missing for most cases. \\ \hline

\end{tabular}
\caption{Scoring Criteria for Completeness and Accuracy.}
\label{tab:criteria}
\end{table}

\end{document}